\documentclass[10pt]{article}

\usepackage{epsf,amsfonts,amssymb,epsfig,amsmath}
\usepackage{multirow}

\usepackage{amssymb,amsmath,mathbbol,mathrsfs}
\usepackage[usenames,dvipsnames]{color}
\usepackage{hyperref}
\usepackage{xcolor}
\usepackage{stmaryrd}
\usepackage{authblk}
\usepackage{framed}
\usepackage{empheq}
\usepackage{slashed}
\usepackage{hyphenat}
\usepackage{cite}

\usepackage{mdframed}

\renewcommand{\text}[1]{#1}

\newcommand{\be}{\begin{equation}}
\newcommand{\ee}{\end{equation}}
\newcommand{\ben}{\begin{displaymath}}
\newcommand{\een}{\end{displaymath}}
\newcommand{\bea}{\begin{eqnarray}}
\newcommand{\eea}{\end{eqnarray}}
\newcommand{\bean}{\begin{eqnarray*}}
\newcommand{\eean}{\end{eqnarray*}}
\newcommand{\ba}{\begin{array}}
\newcommand{\ea}{\end{array}}
\newcommand{\bi}{\begin{itemize}}
\newcommand{\ei}{\end{itemize}}

\newcommand{\bbR}{{\mathbb{R}}}

\def\1f{f_1^{1/2}}
\def\2f{f_2^{1/2}}
\def\4f{f_4^{1/2}}

\begin{document}

\begin{titlepage}

\vfill



\begin{center}
   \baselineskip=16pt
   {\Large\bf Recovering General Relativity from a Planck scale discrete theory of quantum gravity}
  \vskip 1.5cm
Jeremy Butterfield${}^a$, Fay Dowker${}^{b,c}$, \\
     \vskip .6cm
            \begin{small}
      \textit{ ${}^a${Trinity College, Cambridge, CB2 1TQ, UK}\\\vspace{5pt}
       ${}^{b}${Blackett Laboratory, Imperial College, Prince Consort Road, London, SW7 2AZ, UK}\\\vspace{5pt}
		${}^{c}${Perimeter Institute, 31 Caroline Street North, Waterloo ON, N2L 2Y5, Canada}\\\vspace{5pt}
              }
              \end{small}                       \end{center}
\vskip 2cm
\begin{center}
\textbf{Abstract}
\end{center}
\begin{quote}

An argument is presented that if a theory of quantum gravity 
is physically discrete at the Planck scale and the theory 
recovers General Relativity as an approximation, 
then, at the current stage of our knowledge, causal sets must 
arise within the theory, even if they are not
its basis. 

We show in particular that an apparent alternative to causal sets, viz. a certain sort of discrete Lorentzian simplicial complex, cannot recover General Relativistic spacetimes in the appropriately unique way. For it cannot discriminate between Minkowski spacetime and a spacetime with a certain sort of gravitational wave burst.

\end{quote}

\vfill
\end{titlepage}

\tableofcontents

\newpage

\section{Introduction}\label{sec:intro}

Many workers tackling the problem of quantum gravity believe that the
differentiable manifold structure of spacetime in General Relativity (GR) breaks down at the
Planck scale and will be replaced by something else in the eventually successful, deeper theory. 
Some go further and believe that this `something else' has a discrete character. 
The purpose of this paper is not to present directly the case for
discreteness at the Planck scale. Instead, we will argue that, whatever might be the 
ultimate underlying degrees of freedom in a theory of quantum gravity: {\it if} that theory recovers General Relativity as an 
approximation in certain physical circumstances,  {\it and if} it is discrete at the
Planck scale,  {\it then} at the current stage of 
our knowledge, the approach must
produce a {\em causal set}, i.e. a locally finite partial 
order \cite{Bombelli:1987aa}. For, we will argue, whatever might be the nature of the 
ultimate underlying degrees of freedom, a causal set is, so far as we currently know, the only entity that can do the job of being approximate-able by a Lorentzian geometry, while also being discrete at the Planck scale.   

In Section \ref{sec:main}, we state our main argument, in terms of two claims, labelled `Claim 1' and `Claim 2'. Then in Section \ref{sec:causalset}, we develop details about causal sets, so as to justify Claim 1. In Section \ref{sec:lattices}, we support Claim 2, by  showing how a certain sort of simplicial complex (fundamentally discrete and Lorentzian) cannot recover a GR spacetime and argue that the conclusion is more general than the particular example. In Section \ref{sec:loop}, we reply to possible objections. Up until this point, our argument will have been wholly ``kinematical'': that is, setting aside dynamics. So in Section \ref{sec:causal set dynamics}, we briefly discuss dynamics and how it bears on our argument. Then in Section \ref{sec:concl}, we conclude. 

\section{The Argument}\label{sec:main}

We begin by stating and discussing two assumptions that our argument will make (Section \ref{subsec:2asspn}). Then we state our argument, in terms of two main claims (Section \ref{subsec:2claim}). Finally in Section \ref{subsec:causalcentral}, we review the central role of causal order in general relativity. This review forms a backdrop to our defence of the two claims in the succeeding Sections (Claim 1 in Section \ref{sec:causalset}, and Claim 2 in Section \ref{sec:lattices}).

\subsection{Two assumptions}\label{subsec:2asspn}

Let us make two assumptions about a theory 
of quantum gravity, which we call `X'.

{\bf Assumption 1}: In certain physical situations and at large scales, X recovers General Relativity (GR) 
as an approximation. 

{\bf Assumption 2}: X is physically discrete 
at the Planck scale. 

\noindent Neither of these assumptions is precise; and at our current stage of knowledge, they cannot be 
made precise. But most workers will have an intuitive 
feeling for what they mean; and we will now develop them in three extended comments. The first two are about Assumption 1; Assumption 2 comes in to play in the third comment. Note that these comments introduce some jargon we will use. Comment 2 introduces {\em grounding state} for a special sort of state within the theory X. Comment 3 introduces {\em Discrete Physical Data (DPD)} for what a grounding state supplies as the material to which a spacetime in GR is an approximation. 
  \medskip

{\bf Comment 1: Recovering GR}:---  

With Assumption 1, we mean to make a thorough-going commitment to General Relativity as the theory of continuum spacetime and gravity that theory X recovers. In particular, Assumption 1 implies Lorentz symmetry in the recovered theory, where by `Lorentz symmetry', we mean the approximate symmetry enjoyed by General Relativity, sometimes referred to as `local Lorentz symmetry'. In concrete terms, this means that theory X is assumed not to give rise to any of the rival ``modified gravity'' theories  that are characterised by having more than one spacetime metric, or aether fields or even more brutal violations of Lorentz symmetry such as background foliations of spacetime. (A useful recent review of such theories from the perspective of their causal properties is \cite{Carballo-Rubio:2020ttr}.) 

Of course, this is not to say that Assumption 1 means that GR is exactly true. GR is assumed to be an approximation, and theory X will predict deviations from GR which, we hope, will be observable to us, large, late-time observers, in spite of the tininess of the Planck scale. But we intend with Assumption 1 to exclude those deviations being (manifestations of) violations of Lorentz symmetry.  

These remarks mean that we intend Assumption 1 to encompass the view of General Relativity as arising as an effective local field theory.  More precisely, it encompasses such a treatment when that treatment takes all the terms of its effective Lagrangian to be appropriate powers of derivatives of the metric in order to respect Lorentz symmetry (with the cosmological constant problem having been solved somehow).  (This kind of effective Lagrangian for General Relativity is common in such treatments: for example, cf. equations (21) and (46) of \cite{Donoghue:1997} 
and equation (3.1) of \cite{Burgess:2004}.)

We also intend Assumption 1 to include matter degrees of freedom, as well as the spacetime---manifold and Lorentzian metric---degrees of freedom. We  call a  vacuum or non-vacuum solution of the Einstein equations in four dimensions a `solution of GR' or `GR solution' for short.  We will say `spacetime of GR' or `GR spacetime' when we want to talk solely about the component of a GR solution that is the 4-dimensional Lorentzian geometry.

To develop Assumption 1 further, we invoke three familiar examples of the recovery of a physical theory from a more fundamental one. 

General Relativity recovers Newtonian gravity as an approximation
in certain, contingent circumstances. There is much to say---and much has been said---about the details, both technical and philosophical, of GR's recovery of Newtonian gravity, see 
\cite{Havas:1964, Trautman:1965,Malament:1986a, Malament:2012} as a sample from this rich literature. An important part of what this ``recovery'' entails is that  all the data needed to recover, approximately, a non-relativistic spacetime picture with a particular Newtonian gravitational potential satisfying Poisson's equation are present within a particular spacetime in GR.  It is this sufficiency of the physical data in the underlying theory for the recovery of the approximating theory that we will appeal to in applying Assumption 1 to our argument. 

Another example we wish to invoke in developing Assumption 1 is fluid mechanics as an approximation to molecular dynamics in certain situations and at large i.e. macroscopic scales. This is an example pertinent to the combination of Assumption 1 and Assumption 2 in that it involves a continuum approximation. Again, there is much to say---and much has been said---about the details, both technical and philosophical, of continuum approximations to more fundamental, atomic theories of matter, see \cite{chapman:1991, cercignani1987boltzmann} as a sample from this enormous literature. 
Here, we want to emphasise just one point: continuum \textit{approximations} where the number $N$ of fundamental constituents is large but \textit{not} infinite---as opposed to strict continuum \textit{limits}---exist and make sense as emergent from the more basic, more fundamental theories. (Here we understand `emergent' as in  \cite{Butterfield:2011xx, Butterfield:2011yy}.)

Since X is a quantum theory, the understanding of Assumption 1 gleaned from the two examples 
above must be augmented by considering how ordinary quantum 
mechanics can  in certain circumstances be approximated by classical mechanics. Once again, there is much to say---and much has been said---about the details, both technical and philosophical, of quantum mechanics' recovery of classical mechanics; see  \cite{Bell:2004, manyworldsbook, Landsman:2017} for a sample from this enormous literature.  The ongoing disagreements about the foundations of quantum mechanics mean that 
there is here rather {\em more} to say than on the two previous examples.  Indeed, there is currently no consensus on whether or in what way we can understand classical mechanics as having, in fact, been recovered from quantum mechanics.  We will cut through this 
ongoing scientific debate  by taking Assumption 1 to cover both the possibility  that the recovery of GR by theory X is based on a better understanding of  quantum theory than the one we have now, and the possibility that the recovery of GR by theory X relies upon similar interpretational rules of thumb---such as Copenhagen-esque splits, and/or anthropocentric reasoning---to those in use today. 

These three examples of recovery lead one to expect that each of the `physical situations' referred to in Assumption 1, in which a solution of GR is obtained as an approximation, may involve not only: (i) a certain state in the physical state-space of the theory X,\footnote{This state, for all we now know, may well {\em not} be a conventional quantum state such as a vector in---or density operator on---a physical Hilbert space. It may, for example, be one \textit{co-event} from a collection of physically allowed co-events in a path integral-based framework (\cite{Sorkin:2006wq}), or some other concept.} but  also: (ii) a choice of a range of values of certain parameters (or ratios of parameters) defining an approximation scheme---an obvious example being a choice of a relevant physical ``observation'' scale and some concomitant coarse-graining scheme.  Besides, we expect that some physical states in X will correspond to GR spacetimes with singularities, and that such a state---if theory X is to fulfil our expectations of quantum gravity---treats the singularity physically and predictively although the continuum, Lorentzian geometric description breaks down close to the singularity. So a single state in theory X may include both a `physical situation'  away from the singularity that has a continuum  approximation and a `physical situation'---close to and at the singularity---that does not.  Thus we expect a variety of components in the definition of the `physical situations' referred to in Assumption 1. All of that being said, for ease of writing, we will sometimes refer to these `physical situations' as `states'.

\medskip

{\bf Comment 2: Some, but not all, GR solutions}:---\\
 Assumption 1 does not require that {\em all} the 
solutions  allowed by the postulates of (some precise formulation of) General Relativity be recoverable 
from X as approximations. For example, spacetimes with closed causal curves 
 might not be recovered by X. 
 
But crucially, Assumption 2 implies that a GR spacetime can only be recovered from  X if the characteristic distance over which the GR spacetime varies appreciably is everywhere much greater than the Planck length. This condition for spacetimes to be recoverable from X will be part of important conditions which we will introduce later; namely, the {\em discrete-continuum correspondences},  for causal sets in Section \ref{sec:causalset} and 
for simplicial complexes in Section \ref{sec:lorlatt}. So we need to be more precise about its meaning. We say that the characteristic distance over which a GR spacetime, $(M,g)$, {\em varies appreciably is everywhere much greater than the Planck length} iff, for every point $p$ in $(M,g)$ there exists a normal neighbourhood, $U_p$ of $p$ that: \\
\indent \indent (i) is covered by local inertial coordinates centred at $p$, \\
\indent \indent (ii) contains an Alexandrov interval---a causal diamond---that contains $p$ and has  large volume in Planck units; and \\
\indent \indent (iii) is such that everywhere in $U_p$, all the components of the Riemann tensor in the local inertial coordinates are small  in Planck units.\\

 In order to claim that GR is recovered from X, however, a very substantial collection of GR solutions must be recovered from X. This collection certainly includes
all solutions that are currently known to be phenomenologically important
such as large portions of Minkowski spacetime, solutions containing 
black holes, expanding cosmologies and solutions containing 
gravitational waves.  
Here is a very concrete example: there must
be a state in X from which it is
possible to glean data to reconstruct, approximately,
a GR solution including gravitational waves propagating from a binary black hole merger event to an observation event
that coordinates with the  LIGO data for GW150914 \cite{PhysRevLett.116.061102}. To put the point contrapositively: if there is no state in theory X that
gives data to which a GR solution with gravitational 
waves is a continuum approximation, then GR is not recoverable from X. 

Let us refer to the states in X that give rise to data from which solutions of GR can be recovered
as   {\bf grounding states}: since they are the ground or basis of recovering GR solutions. (We avoid: (i) ``semiclassical states'', as having perhaps, too specific connotations; and (ii) ``basic states'', as connoting that the states are fundamental within X---whereas we want to allow that the grounding states might well be specified using criteria that are {\em not} fundamental in  X.) One can conceive of the whole collection of grounding states in X as  the \textit{continuum regime} or \textit{continuum sector} of X.  Comment 3 will fill out the idea of a grounding state, by developing Assumption 2. \\

To conclude Comments 1 and 2: we of course admit that  Assumption 1 is not {\em compulsory}. There is however, to date, no convincing observational evidence for any theory of gravity apart from GR. And so we expect widespread, though not unanimous, assent to Assumption 1. \\

{\bf Comment 3: From discrete data to one spacetime}:---\\
For our argument,  we take Assumptions 1 and 2 as implying that: 
\begin{quote}
 {\it a
grounding state of X contains, or gives rise to some Discrete Physical 
Data (DPD),\footnote{\label{Riemann}{An alternative phrase that we might have adopted instead of `DPD' is Riemann's phrase \textit{Discrete Manifold} in his  {\em Habilitationschrift} (\cite{Riemann:1868}, see \cite{Riemanntranslation:2016} for an English translation) where
`manifold' means simply `multitude'. But owing to the development of Riemannian geometry (which of course Riemann's lecture began), the term `manifold' has come to 
 so strongly connote the continuum that `discrete manifold' can sound to modern ears like an oxymoron.}} to which a GR spacetime is an---essentially unique---approximation.}
\end{quote} 
This Comment spells out the meaning of this italicized statement, in three Remarks. The first is about DPD, the second about essential uniqueness; and the third is about discreteness and about the Planck scale. 

\medskip

\noindent {\em  A: The grounding state and the DPD}:  
 In what follows, we will be mostly interested in the data to which the GR spacetime is an approximation, and will refer to it as {\em the DPD} in the grounding state. Agreed: if the grounding state recovers a non-vacuum GR solution then that grounding state will produce, in addition to the DPD to which the GR spacetime is an approximation, other discrete data that encode the matter distribution of the GR solution. So if we need to refer to the additional matter data,  we will do so explicitly, in addition to the DPD. 

We do not specify how the grounding state contains, or produces, or gives rise to the 
DPD: for example, this data could be eigenvalues of or expectation values of certain operators on a Hilbert space---or not. 
Whatever theory X is, if it recovers GR as an approximation, there
must come a stage  at which we have in our hands a set of DPD and there is a 
procedure, in principle, for obtaining a
GR spacetime, approximately and essentially uniquely 
from that set of DPD. This stage is
{\it after} the quantal nature of the grounding state---small scale quantum fluctuations,
interference between histories, multitudes of branches or what have you---has been ``dealt with'' by coarse-graining, collapse of the wave-function, anthropocentric reasoning or what have you. 
In the end, if the theory is to work for {\it this} purpose---recovering a GR spacetime as an approximation in some state---then
the data it produces must be data one can store as
bits in a classical computer memory.

\medskip

\noindent  {\em  B: Essential uniqueness}: 
By ``essentially uniquely''  we mean that if two continuum
spacetimes $(M,g)$ and $(M', g')$ are both recoverable from the DPD provided by a grounding state in X, then they must be approximately isometric on scales large compared to the Planck scale. 

There are two concerns that one might have about this requirement. Broadly speaking, (i) is mathematical, while (ii) is physical. 

\medskip

(i):  One might worry about the formal existence of a scale-dependent metric on the space of Lorentzian geometries that would allow us to say when two geometries are indistinguishable above a certain scale. We will mention one proposed scale-dependent distance function on Lorentzian geometries in our discussion of causal sets. However, the concept of spacetimes being `approximately isometric above some scale' will make intuitive sense to most workers on quantum gravity---cf. Kaluza-Klein theory---and we rely on this heuristic understanding being sufficient for our argument.

\medskip

(ii): One might worry that  orthodox quantum theory (on at least some views of it) licences superpositions of macroscopically very different mass distributions: which in the context of GR, would seem to lead to superpositions of spacetimes that are {\em not} approximately isometric. Indeed, maybe the grounding states in theory X are superpositions of this kind. 
Our answer to this has already been given in Remark A. Namely, our assumption that theory X recovers GR includes the assumption that such superpositions, if they are in fact present, can be dealt with, and have been dealt with by the stage at which we consider the DPD to which a spacetime in GR is a good approximation.

One might also ask: `what about dualities?'
In recent decades, fundamental physics has formulated---and in some cases proven---various dualities between quantum theories formulated on spacetimes. A duality is in effect a surprising isomorphism between the state-spaces and algebras of quantities of the two theories, like a giant symmetry between the theories \cite{Deharo:2017, Deharo:2019}. Indeed, in the most celebrated case, \textit{viz.} gauge-gravity duality in string theory, the two spacetimes concerned even differ in their dimension, one being the boundary of the other: which certainly counts as surprising! So if we countenance dualities such as these, what becomes of our `essential uniqueness' requirement?

We answer this concern in the same way as the concern about quantum superpositions: they are important concerns, \textit{but they are not our concerns, here}.  In GR, the physical world is just one spacetime (and matter distribution). If theory X enjoys a duality or dualities, and if it recovers GR, then this singleness of spacetime must be recovered, somehow, in X. We step in and take the DPD in our hands \textit{after} this singleness has been established,  however it has been established: i.e. established by whatever argument disposes satisfactorily of all but one of the dual descriptions. (It could for example be a more-or-less anthropocentric argument.)

\medskip

\noindent {\em C: Planck scale discreteness}
For definiteness let us define\footnote{We use 
$8\pi G$---the cosmologist's convention---rather than just $G$ in the definitions} the Planck time, Planck length and Planck 4-volume, respectively:
\begin{equation}
t_{p} : = \sqrt{\frac{8 \pi G \hbar}{c^5}}\,,\quad
l_{p} : = \sqrt{\frac{8 \pi G \hbar}{c^3}}\quad \textrm{and}\quad 
V_{p} : = t_{p}l_{p}^3 = (8 \pi G \hbar)^2 c^{-7}\,,
\end{equation}
where $G$ is Newton's gravitational constant, $c$ is the speed of light and $\hbar$ is the (reduced) Planck 
constant. 
In GR, $V_{p}$, $t_{p}$ and $l_{p}$ are not independent: knowing one of the three quantities above is sufficient to fix the other two.

At our current stage of knowledge of quantum gravity based on work in any existing approach, 
we cannot say more than that in our putative successful theory X the discreteness time scale, say, will be \textit{of order} $t_{p}$.   However, Assumptions 1 and 2 imply that in our theory X that successfully recovers GR, the actual value of the discreteness scale should be calculable. In other words, theory X should tell us, via
the discrete-continuum correspondence, the ratio of the fundamental discreteness time scale $t_{f}$, say, and $t_{p}$:
\begin{equation}
t_{f} = z t_{p},  \quad l_{f} = z l_{p}, \quad V_{f} = z^4 V_{p}\,,
\end{equation}
where $z$ is a number of order 1 that theory X should determine. 
One way  $z$  might be determined is by calculating the entropy
of a black hole in theory X in terms of the fundamental discreteness length: $ S_{BH} =\alpha \frac{A_{BH}}{l_{f}^2}$. Equating this to the known value---$S_{BH} =2 \pi \frac{A_{BH}}{l_p^2}$---would then give the value of $z = \sqrt{\frac{\alpha}{2 \pi}}$. 

The condition that theory X is physically discrete at the Planck scale
has meaning only for the grounding states. For it is only these states (by definition) 
that provide DPD that can be approximated by
Lorentzian geometries.   So,  for a grounding state---and only for a grounding 
state---the DPD are approximated by a 
GR spacetime so that the \textit{discrete-continuum correspondence} between the set of DPD and the spacetime does justice to the concept of Planck scale discreteness. 

We do not specify exactly how the DPD are discrete. 
One might want to demand of theory X that the geometry of a GR spacetime 
in a region of finite spacetime volume be recovered from a finite amount of combinatorial data---as 
in a causal set---but our argument will not hinge on it. The physical data provided by 
theory X in a grounding state may be 
completely digital and combinatorial, or they may include real numbers such as edge lengths on a graph. 

But we do take Assumption 2 to demand that the DPD and its approximating spacetime together
embody the two heuristics: ``to fully describe continuum physics in and of a 
 finite region in a GR spacetime requires only a finite amount of information'' and ``the DPD do not give information about geometry on scales smaller than the Planck scale.''

For example, a lattice whose spacing is many orders of magnitude smaller
than the Planck length cannot arise as DPD in a grounding state of X. 
Nor can the  {\it geometrical} form of a simplicial complex as a piecewise flat continuum manifold---a union of pieces of flat Minkowski space---arise as DPD in X. The flat ``filling'' of the interiors of the simplices is continuum information. 
A simplicial complex is only discrete if it is a {\it{combinatorial}}
complex, perhaps decorated with edge lengths, triangle areas, spins etc. This exclusion will be important in Section \ref{sec:lattices}'s defence of Claim 2.

The scale of the discreteness of theory X is presented in this paper as discreteness in spacetime. We briefly consider what it would 
mean for this discreteness to be expressed in terms of energy, because some workers conceive of the `Planck scale' in the first instance as very large (in energy) and not, primarily, as very small (in spacetime). 
Discreteness at the Planck scale means that there is a physical cutoff in spacetime which one might want to translate directly into a high frequency cutoff at the Planck frequency, $\nu_p: = t_p^{-1}$, and, since frequency is a frame-dependent concept, one might think that a high frequency cutoff must imply a preferred frame in which to express that cutoff and thereby violate Lorentz symmetry. 
The relationship between spacetime discreteness, a high frequency/energy cutoff and Lorentz symmetry
 is more subtle than this suggests, however, and in particular the phrase `frequency is a frame-dependent concept' is 
a little too glib in this context. Frequency \textit{is} a frame dependent concept \textit{in the sense} that a free wave packet in Minkowski spacetime 
whose frequency is peaked around  $\nu< \nu_p$ in one frame is peaked at a different frequency, $\nu'$,  in another frame. Depending on the boost factor, $\nu'$ can be greater than $\nu_p$. 
 The point is that such a packet really has no associated covariant energy,  except the rest mass if it is a massive particle/field. Thinking about this in covariant, spacetime terms, one sees that the crucial question for compatibility with discreteness is whether the \textit{spacetime support} of the packet corresponds to a large enough subset of the DPD to recover all the contours of the packet. This \textit{will} be the case if there exists a frame in which the typical frequency of the packet is small compared to 
$\nu_{p} $ and if the discreteness of the DPD is Planck scale, and Lorentz invariant. Since,  then, in every frame \textit{including the one in which the packet happens to have low frequency} there will be enough DPD corresponding to the 
spacetime support to carry the information about the contours of the packet  (Section 1 of \cite{Sorkin:2009bp}).

Lorentz invariant discreteness, then, is compatible with free wave packets of any frequency in any frame in infinite Minkowski spacetime; 
and consideration of these packets
does not really probe the concept of a high energy cutoff (except in the trivial sense 
that the rest mass of an elementary particle should be less than the Planck mass---for, otherwise, the frequency, 
$(2\pi)^{-1} \sqrt{m^2 + |\vec{p}|^2}$,  is Planckian or higher in \textit{every} frame).  It is in considering 
\textit{interactions} that the concept of `high frequency cutoff' becomes meaningful because 
there \textit{are} covariant, frame-independent measures of the energy of an interaction. For example, the centre of mass energy of a collision of two particles is Lorentz invariant. Again, translating this into a covariant spacetime picture, the  question becomes: are there 
enough DPD corresponding to the interaction region---the intersection of the spacetime supports of the two
wave packets, say---to support the physical details of the interaction? Now, if the centre of mass energy of a two particle collision 
 is higher than the Planck energy and the spacetime volume of the interaction region is smaller than the Planck volume,  there will
 be very few or even no corresponding DPD to support the interaction, effectively cutting off such high energy interactions.\footnote{In the context of causal set theory, Sorkin has pointed out that this is a potential signature of discreteness: such packets propagating on a 
 causal set background could pass through each other without interaction \cite{Sorkin:2009bp}.} In this way,  a Lorentz invariant 
 high energy cutoff arises from Lorentz invariant spacetime discreteness.

\subsection{Two claims}\label{subsec:2claim}
As announced in Section \ref{sec:intro}, we now state our main argument, in terms of two claims, `Claim 1' and `Claim 2'. Then we will make explicit two limitations of our argument, before turning to support the Claims: Claim 1 in Section \ref{sec:causalset}, and Claim 2 in Section \ref{sec:lattices}.
\bigskip

{\bf Claim 1}: {\sl A causal set---a locally finite partial 
order---is a set of DPD that, taken as being discrete on the Planck scale, can recover a 
GR spacetime  as a continuum approximation.}\\

\bigskip 
  
{\bf Claim 2}: {\sl There is in the current literature no other proposal for a set of Planck scale DPD that can recover a GR spacetime as a continuum approximation.}

\bigskip

Thus Claims 1 and 2, as clarified by Assumptions 1 and 2 in Section \ref{subsec:2asspn},  express the argument we announced in Section \ref{sec:intro}. Namely: whatever the 
ultimate underlying degrees of freedom in a theory X of quantum gravity, {\it if} X is discrete at the
Planck scale, and recovers General Relativity as an 
approximation in certain circumstances,  {\it then} according to our current state of knowledge, X needs to produce a  causal set. 

But before we support the Claims in Sections \ref{sec:causalset} and \ref{sec:lattices}, we should clarify that our argument has two important limitations of scope.  

\bigskip

{\em A: We only consider kinematics:} We  stress that Claim 1 does not say that {\em every} causal set recovers a GR spacetime. Only some do; and  which ones do  will be spelled out in Section \ref{sec:causalset}. For the Claim is made at what one might call a ``kinematical level''.  At the current stage of our knowledge, no theory of quantum gravity, neither causal set theory nor any other theory explains dynamically why grounding states with DPD that can recover GR spacetimes occur. In the case of causal set theory,  finding a quantum dynamics that will do this job is a major outstanding task. We will say a little more about this at the start of Section \ref{sec:causalset}, and in Sections \ref{subsubsec:agnm} and \ref{sec:causal set dynamics}. 

\bigskip

{\em B: We target only theories that produce discrete physical data:} 
 There are theories, and frameworks for theories, of quantum gravity that \textit{use} discreteness,  but that use in itself does not 
bring them into the scope of our argument, for one or both of two reasons. (i): The first reason is that the theory uses discreteness either as a regulator to be taken away in a strict continuum limit, or more generally to define an approximation to a continuum theory. (ii): The second reason is that, even away from the so-called continuum limit, at a stage at which the regulator is still finite (non-zero), the spacetimes considered are actually continuous because they are \textit{geometric} simplicial complexes---piecewise flat Lorentzian manifolds---of the kind discussed in Remark C of Comment 3 in Section \ref{subsec:2asspn}. 

Both these points (i) and (ii) are well illustrated by the construction of the path integral in elementary non-relativistic quantum mechanics. Thus recall that in the process of defining the path integral, we consider zigzag (i.e. piecewise straight) paths in spacetime with successively shorter time divisions, so that at each finite stage of the construction all the paths considered (`skeletonisations') are everywhere continuous but not differentiable at their corners. Taking an appropriate limit of infinitely many time divisions, we get the path integral, with all its new mathematical structures and physical ideas. (So this construction exemplifies the idea of a singular limit, in the---non-technical---sense of a limit in which some facts or structures, mathematical and/or physical, exist at the limit, but not before: as mentioned in \cite{Butterfield:2011xx, Butterfield:2011yy}.)  Though one might casually say that the skeletonisations, being zigzags, ``look discrete'', that is obviously loose talk, since they are continuous paths. Moreover, these zigzag paths are just stages in the construction, and none of them (no matter how fine, i.e. how minuscule the time duration of their straight segments) give the physical interpretation of the path integral eventually constructed. Thus our point here is that, for all its novelty and richness---its glories!---the use of a form of discreteness in defining the path integral in quantum mechanics does not in itself produce discrete physics. It is wrong to say that the path integral in quantum mechanics is discrete.\\

Similarly here: theories such as Causal Dynamical Triangulations (CDT) \cite{Ambjorn:1998xu,
Ambjorn:2000dv,Ambjorn:2000dj,Ambjorn:2001cv} and 
Lorentzian Quantum Regge Calculus \cite{Williams:1986} are examples of 
quantum gravity approaches that use piecewise flat Lorentzian manifolds.  CDT uses a path integral approach in which the piecewise flat Lorentzian manifolds play the role
of the skeletonised trajectories in the quantum mechanics example, and the full CDT theory is defined in the  limit in which the
lengths of the edges of the simplices in the (Euclideanised) piecewise flat manifold are taken strictly to zero. 
Such use of piecewise flat Lorentzian manifolds in quantum gravity theories and frameworks is {\em not} targeted by our argument, in particular by Claim 2.  Nor does our argument directly target the use of combinatorial Lorentzian simplicial complexes \textit{if} the discreteness scale is
not physical but is a regulator that is taken strictly to zero in the definition of the quantum gravity theory.

Now, there are several quantum gravity approaches---in addition to CDT and Quantum Regge Calculus which we have mentioned---that have a discrete flavour, including: Energetic Causal Sets \cite{Cort_s_2014}, Entropic Gravity \cite{Verlinde_2011}, Group Field Theory \cite{Oriti:2009}, Holography  \cite{10.3389/fphy.2020.00111}, Loop Quantum Gravity \cite{rovelli:2008},  Spin Foams \cite{rovelli_vidotto_2014},  Quantum Graphity \cite{Konopka_2008}, Spacetime Code \cite{Finkelstein:1969}, Thermodynamic Gravity \cite{Padmanabhan:2014}, and the Wolfram model \cite{Gorard_2020}, among others. It would be a good project to assess these discretely flavoured quantum gravity approaches as to whether they only employ discreteness {\em \`{a} la} (i) and-or (ii) above, or whether they aim to recover a GR spacetime in Section 2.1's sense by means of Planck scale discrete physical data and thereby fall within the scope of our argument. Any quantum gravity approach within our scope is then in danger of stumbling in the way that Section \ref{sec:lattices} will show combinatorial Lorentzian simplicial complexes do. 

\subsection{General relativity and Lorentzian geometry}\label{subsec:causalcentral}

Our arguments in this paper depend on the all-important discrete-continuum correspondence, i.e. the correspondence between a set of DPD  and the continuum spacetime that approximates it. To analyse and assess this correspondence, we obviously need to understand well what it is that the DPD in a grounding state of theory X must recover: in short, a GR spacetime. 
So in this Subsection, we will briefly review GR spacetimes, in the sense we adopted in Comment 1 of Section \ref{subsec:2asspn}: that is, as Lorentzian manifolds $(M,g)$.

A Lorentzian manifold---in stark contrast to a Riemannian manifold---is not isotropic on small scales because Minkowski space itself, though flat, is not isotropic: it has preferred directions, the \textit{null} directions. Moreover, the points along a null geodesic are totally ordered in the spacetime causal order\footnote{This use of the word `causal' in GR does not imply `causation' (whatever that might mean!): to say $x$ is to the causal past of $y$ is merely to say that $x$ is before $y$ and $y$ is after $x$.}: it is meaningful to say that point $a$ is to the past of point $b$ on a null geodesic in Minkowski spacetime even though the distance between them is zero. 
This remarkable structure---so very different from Riemannian geometry---means that Lorentzian geometry, if it is not actually nonlocal, teeters on the very edge of being nonlocal. This verging-on-nonlocal character can be illustrated in the following way. Consider a point $p$ in 4-dimensional Minkowski spacetime. All the points along the future and past light cones from $p$ are zero distance from $p$. And the locus of points that are one Planck unit of geodesic proper time, say, to the past or to the future of $p$ is a double-sheeted 3-dimensional spacelike hyperboloid of infinite volume that asymptotes to the past and future light cones from $p$. So the set of points that are physically, geometrically close to $p$ is very nonlocal. Besides, one can also consider mathematically the locus of points that are one Planck unit of spacelike geodesic distance from $p$: a 3-dimensional single-sheeted hyperboloid that asymptotes to both the past and future light cones. Putting these remarks together means that the closest analogue of a Euclidean geodesic ball around a point $p$ is an unbounded 4-dimensional region of infinite spacetime volume between these hyperboloids, a neighbourhood of the past and future light cones from $p$ as shown in Figure 1.

\begin{figure}[h!]\label{figure1} 
\centering
{\includegraphics[scale=0.5]{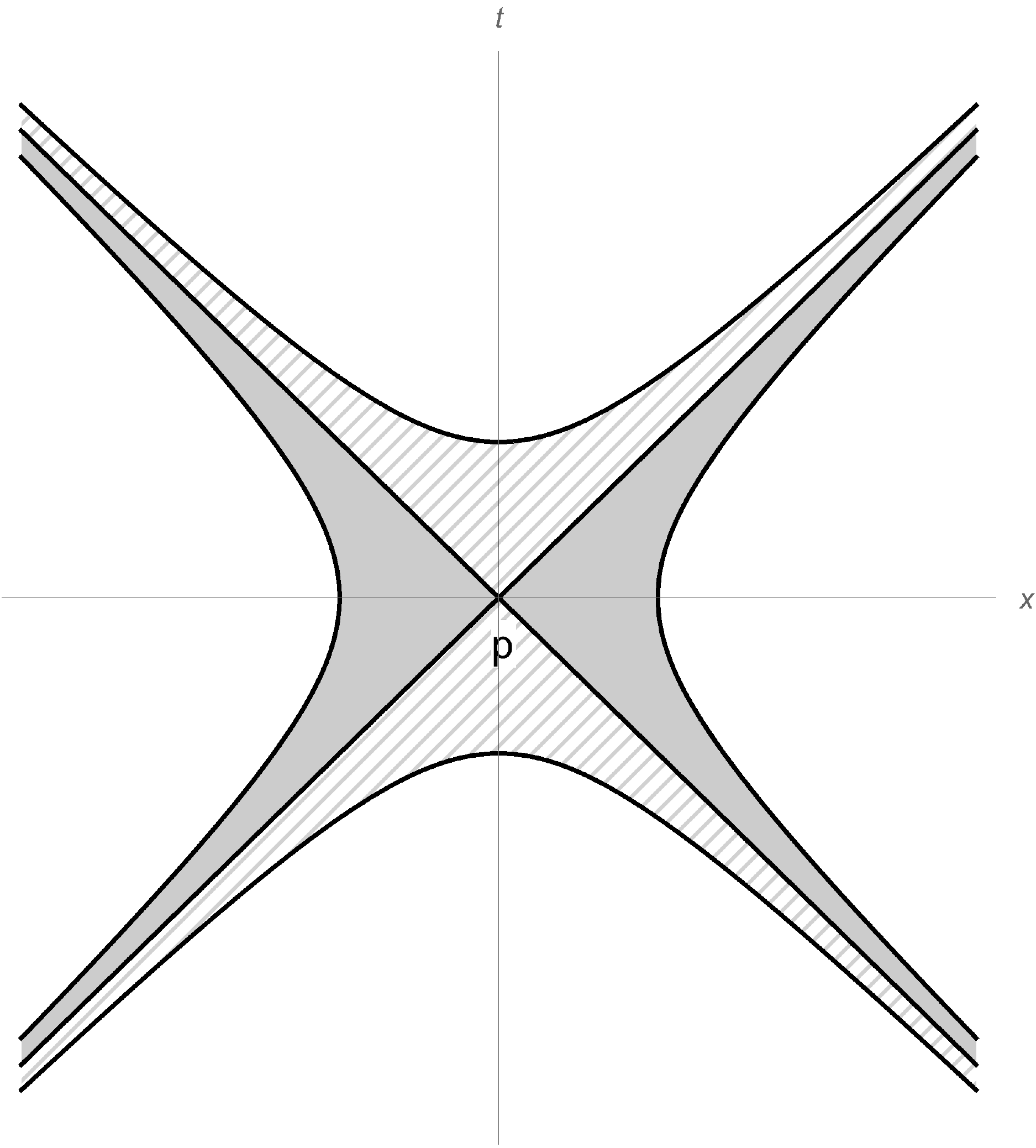}}
\caption{The figure shows the $t-z$ plane of 4- dimensional Minkowski spacetime. The figure should be 
imagined as rotated about the $t$-axis into the suppressed $x$ and $y$ dimensions. The point $p$ is at the origin and the light cones of $p$ are shown. The striped shaded region is the locus of points in the causal past or causal future of $p$ 
that are within one unit of geodesic proper time of $p$. The solid shaded region is the locus of points spacelike to $p$ that are within one unit of geodesic proper spatial distance of $p$. Both shaded regions are unbounded and of infinite spacetime volume and their  union is the (analogue of the) geodesic ball around  $p$. The boundaries of this ball are hyperboloids. }
\end{figure}

The nonlocal character of these Lorentzian ``geodesic balls'' immediately calls into question the relationship between the metric and the topology of Lorentzian geometries. 
In Euclidean space,  the geometry and the topology are compatible in the sense that the set of open geodesic balls is a base for the manifold topology, and this is also the case 
for Riemannian geometries generally. In Minkowski spacetime, the nonlocality of the geodesic ``balls'' means they cannot form a base for the topology. Remarkably, the compatibility between geometry and topology nevertheless holds for Minkowski spacetime, not via geodesic balls at points but, instead, via Alexandrov intervals---``causal diamonds''---between 
\textit{pairs} of points. The Alexandrov interval between points $x$ and $y$, where $x\in I^-(y)$ (i.e. there is a past directed timelike curve from $y$ to $x$), is the intersection of the chronological future $I^+(x)$ of $x$ (i.e. the set of points to which there is a future directed timelike curve from $x$) and the chronological past  $I^-(y)$ of $y$. The Alexandrov topology is the smallest topology containing these Alexandrov intervals and the set of Alexandrov intervals is a base for the Alexandrov topology. For Minkowski spacetime, the Alexandrov topology is equal to the manifold topology. This result continues to hold for other Lorentzian geometries, but---in contrast to the Riemannian case---not for all of them: the Alexandrov topology equals the manifold topology if and only if the Lorentzian geometry is strongly causal (page 487 of \cite{Kronheimer:1967}).\footnote{Strong causality is a weaker causality condition than global hyperbolicity but a stronger condition than being merely causal: see \cite{Minguzzi:2006sa} for a review of the rungs of the ``causal ladder''.}  Note the difference between the Riemannian and Lorentzian cases. In the Riemannian case, the metric determines the topology locally and via the concept of geodesic distance.  In the Lorentzian case the metric determines the topology bi-locally and via the concept of causal order.\footnote{Recall that the causal order is the totality of relations $x\in J^-(y)$ (which means there is a future directed causal curve  from $x$ to $y$). For any Lorentzian spacetime, the causal order determines the chronological order and thence the Alexandrov topology (page 485-487 of \cite{Kronheimer:1967}).}

Remarkably, the causal order of a strongly causal Lorentzian manifold determines much more than just its topology. It also determines the differentiable structure in dimensions greater than 2, thanks to a 1966 theorem of Hawking \cite{HawkingAdamsPrizePublished, Hawking:1976fe}. With the tangent space in hand, the metric can then be determined in terms of one unknown fiducial component, which component can be traded in for an overall conformal factor (cf. pages 60 and 61 in  \cite{HawkingEllis}).  Thus, for a strongly causal spacetime of dimension greater than 2, the causal order determines the metric up to an overall conformal factor. 
This result was later extended by Malament to cover Lorentzian manifolds satisfying a strictly weaker causality condition \cite{Malament:1977}. This final result, the Kronheimer-Penrose-Hawking-Malament (KPHM) theorem, states that the causal order of a distinguishing\footnote{A past/future distinguishing spacetime is one in which two distinct points have 
distinct chronological pasts/futures. A distinguishing spacetime is both past and future distinguishing.}   
spacetime of dimension greater than 2 determines its topology, differentiable structure and 
metric up to a conformal factor. It is the strongest statement of how much of the structure of a Lorentzian manifold is determined by  the causal order.\footnote{Here, `the strongest  statement' means that any weaker causality condition renders the theorem false.} In short: the causal order captures almost the full geometry and all that is missing is a local scale. The KPHM theorem can be encapsulated in the  slogan: ``Order + Volume = Lorentzian Geometry.'' There is no Riemannian analogue. 

Causal order is not just at the heart of the mathematics of Lorentzian geometry in GR, it is also at the heart of the physics of GR. Carter-Penrose diagrams and other spacetime diagrams representing the causal structure of spacetimes abound in GR textbooks. The epitome of GR is a black hole and the event horizon of a black hole is defined in causal terms: the boundary of the closure of the causal past of future null infinity. From this concept of event horizon flows much of the physics of black holes. For example: the Second Law of black hole mechanics, Hawking's Area Theorem, is a result about the event horizon \cite{PhysRevLett.26.1344}; and assuming an event horizon forms, it can be proved that the Kerr family of solutions is very likely to be the physical description of the final state of any black hole in GR (either with the assumption the final state is axisymmetric \cite{PhysRevLett.26.331,PhysRevLett.34.905} or 
dropping the assumption of axisymmetry but assuming that the metric is analytic (Section 9.3 of \cite{HawkingEllis}, Theorem 33 of \cite{chrusciel:2012})). 

``Gravity is geometry" is a common slogan to sum up the lesson of GR. But a \textit{Riemannian} curved spacetime would not 
give us the physics of GR.  Perhaps we should always say ``Gravity is Lorentzian geometry" to remind ourselves of what GR is really teaching us. 

\section{Causal Sets}\label{sec:causalset}

In this Section, we will support Claim 1 by referring to existing results in the literature on causal sets. Indeed the central points we make are present already in the founding papers of causal set theory \cite{Myrheim:1978, tHooft:1979, Bombelli:1987aa}, though more evidence for them has accrued since then. 

So the Claim we wish to defend is:--- Given any distinguishing GR spacetime, $(M,g)$, whose geometry varies appreciably only on scales much larger than the Planck scale, there is a causal set $C$ that is: 
 \begin{itemize}
\item[(i)]  well approximated by $(M,g)$,  and
\item[ (ii)] such that the correspondence between  $C$ and $(M,g)$ is
discrete at the Planck-scale. 
\end{itemize}
More precisely: there are many causal sets that are very likely to satisfy (i) and (ii). Here, of course, `very likely' signals, not a probability, but our having good theoretical reasons to believe this. This use of `very likely' echoes the distinction we made in Remark A at the end of Section \ref{subsec:2claim}, between kinematical and dynamical aspects of causal set theory. That is: we take Claim 1 as a kinematical claim, about the approximation relation between a GR spacetime and a causal set. So our argument does not depend on assumptions or results about the dynamics of causal sets:  which, admittedly, is a topic that raises important questions that we will briefly discuss in Sections \ref{subsubsec:agnm} and \ref{sec:causal set dynamics}.

As we will explain, there {\em are} issues to address in causal sets' satisfying (i) and (ii): issues that have been recognised since the inception of the causal set approach to quantum gravity. Describing these issues, and the progress made so far in addressing them, will be the main theme of this Section. We begin with a warning that the discrete-continuum correspondence---the  approximation relation---is not as straightforward as one might think  (Section \ref{subsec:warn}). 
In Section \ref{subsec:kin}, we give the all-important formulation of the discrete-continuum correspondence for causal sets. Then the role of Poisson sprinkling is explained in Section \ref{subsec:poisson}.
Then in Section \ref{subsec:evidence}, we describe some of the evidence produced so far for our Claim about causal sets' ability to recover GR spacetimes.

\subsection{Approximating the discrete by the continuous}\label{subsec:warn}

\subsubsection{A variety of analogies}\label{subsubsec:analogies}

In Comment 1 in Section \ref{subsec:2asspn}, we gave fluid mechanics as an example of an emergent continuum from a discrete substructure. Bulk matter provides other examples, including metals and other crystals that are formed from regular arrays of discrete atoms. Everyday life provides yet more examples.  We are all familiar with how a digital photograph, with its finitely many pixels, can be indistinguishable to us humans from the photographed scene, which is---so one presumes, in a classical world!---an analogue scene: i.e. a scene whose exact description requires continuously many real numbers.\footnote{Even for black-and-white photography, the intensity, i.e. brightness, would be, according to a classical electromagnetic description,  a continuous field, i.e. a map from physical space $\bbR^3$  to $\bbR$.} A film reel of a  sequence of finitely many still images is experienced by the watcher as continuous motion. A geodesic dome formed of many flat triangles appears, when viewed from far enough away, as a smooth surface.  

These examples and experiences give rise to intuitions about emergent continua that may tempt us into thinking that it is bound to be straightforward to recover a GR spacetime from a set of DPD that is discrete at the Planck scale. `Straightforward' because such examples abound and because the Planck scale is so very very small---not only compared to us humans but also compared to the scale of physics we have so far probed in any particle accelerator---so that \textit{of course} a discrete manifold can appear as a  continuous spacetime to us. 

We want to endorse the idea that a continuum spacetime can indeed emerge from a discrete underpinning.  But at the same time, we {\em caution} against the thoughts that it is bound to be straightforward, and that we can take {\em any} of the concrete examples of emergent continua mentioned above as a guide to the relation 
between the DPD and the GR spacetime. For we will argue, in this Section on causal sets and the next on simplicial complexes, that the approximate Lorentz symmetry of GR and our assumption that the DPD are discrete at the Planck scale mean that only some of these examples are good guides to the discrete-continuum correspondence by which theory X recovers GR.  In particular, it will become clear in Sections \ref{subsec:kin}, \ref{subsec:poisson} and \ref{subsec:evidence} that the molecule/fluid analogy is a fruitful one, whereas an atom/crystal analogy is {\em not}. And in Section \ref{sec:lattices}, we will see that a natural discrete-continuum correspondence for simplicial complexes {\em fails}.

\subsubsection{Agnosticism about the existence of structure ``below'' the Planck scale}\label{subsubsec:agnm}

Recall how, early in our mathematical education, points are introduced as being the limit of ever smaller regions of space or spacetime. The teacher or textbook urges on us that there is (``surely'') no lower bound to the diameter or volume of these regions; and we come to accept points as extensionless in the above sense. (Hence the label, in pedagogy about spacetime, `point-events'.) Assumption 2, of Planck scale physical discreteness, is that what the teacher or textbook urged is wrong in quantum gravity. The assumption denies the physical meaningfulness of any continuum topological or geometrical concepts such as dimension,  length and volume below the Planck scale in a GR spacetime. 

But this is not to say that there cannot be any physical discrete structure \textit{in addition} to the DPD and to data that recover any matter degrees of freedom in the GR solution. One might be tempted to call such extra structure \textit{sub-Planckian} but that could be misleading as such structure really has no scale as such because it does not contribute to the recovery of the approximating Lorentzian geometry $(M,g)$. 
To illustrate this idea of extra structure, consider the case of causal sets. We are about to argue (in Section \ref{subsec:kin}) that, using this paper's jargon of grounding states and DPD,  a causal set can be the Planck scale DPD in a grounding state that recovers a GR spacetime. The elements of such a causal-set-as-DPD, $C$, are to be considered as {\em atoms of spacetime} in the sense of the original Greek word {\em atomos}, i.e. indivisible. That is: an element of $C$ has no internal structure and cannot be analysed (`divided') into parts. But that is not to say that there can be no physical data at all in addition to $C$: it is just that that extra data will not be data about the approximating Lorentzian geometry. 

Now, if there \textit{are} further data as well as the DPD in the grounding state, then who knows what they are and who knows how the grounding state contains, or implies, or gives rise to\footnote{Here, the phrase `contains, or implies, or gives rise to' deliberately echoes the italicized statement at the start of Comment 3, Section \ref{subsec:2asspn}. } the DPD themselves. A certain amount of some sort of coarse-graining will have had to have been done to the full discrete data provided by the grounding state in order to arrive at the DPD alone. In the specific case of the causal-set-as-DPD, $C$, from which the GR $(M,g)$ spacetime is recovered, one possibility is that the grounding state might provide a causal set $C'$ that is not itself faithfully embeddable in a GR spacetime but which, on decimation of $C'$ by a random deletion of elements chosen with fixed probability, produces the DPD-set $C$ that does faithfully embed in $(M,g)$ at Planckian density.

\subsection{The discrete-continuum correspondence for causal sets }\label{subsec:kin}   

In Section \ref{subsec:causalcentral} we reviewed key aspects of  Lorentzian manifolds, the GR spacetimes that must be recovered by grounding states in theory X. 
Consider, now, a \textit{particular}  4-dimensional, distinguishing GR spacetime $(M,g)$ that must be recovered in quantum gravity theory X. In this Section we will state the discrete-continuum correspondence---or approximation relation--- between a causal set as Planck scale DPD\footnote{When we say `the set of  DPD is a causal set', we mean, strictly speaking, an order-isomorphism equivalence class of 
causal sets. However, we adopt the standard practice whereby we think of and talk about the set of DPD as a single representative causal set rather an equivalence class of causal sets; and we then compensate for this by making sure that the mathematical identity of, and any labels on, the elements of the causal set  are ignored in the causal set's role as DPD, so that only the \textit{order relation}, $\prec$, between the elements and the \textit{number} of the elements have meaning as physical data.} 
 and $(M,g)$ \cite{Bombelli:1987aa}. But first, for completeness' sake, we give the formal definition of a causal set.

 \emph{Definition}:  A {\em causal set} is  a set $C$ with a relation $\prec$, called ``precedes'', on $C$ that satisfies the following conditions: 
\begin{itemize}
\item[(1)] if $x \prec y$ and $y \prec z$ then $x \prec z$ \,
$\forall x,y,z \in C$ (transitivity); 
\item[(2)] $x \nprec x$\, $\forall x \in C$ (acyclicity);
\item[(3)] for any pair of elements $x$ and $z$ of $C$ such that $x\prec z$, the set
$\{y | x \prec y \prec z \}$  is finite (local finiteness).
\end{itemize}
Of these axioms, (1) and (2) together say that $(C, \prec)$ is a partially ordered
set, or poset for short. A poset is sometimes called simply an `order'. 
The third axiom of local finiteness expresses the discreteness of the causal set . 
Note that the continuum spacetime causal order is reflexive: $p \in J^-(p), \, \forall  p\in M$ so it is the relation `precedes or equals' in the causal set, with notation 
$\preceq$,  that coordinates with the continuum spacetime causal order in the discrete-continuum correspondence below. An alternative convention---`the reflexive convention'---in the 
literature on partial orders is to give the definition in terms of $\preceq$.

 Phrases that can be taken as synonymous with causal set are ``discrete order'', ``locally finite partial order'' and ``transitive directed acyclic graph."

Now let  {\em the discrete-continuum correspondence for causal sets}, which we will call {\bf DCC-C}, be as follows:

\medskip

\begin{mdframed}
A  causal set, $(C, \prec)$,  recovers the GR Spacetime $(M,g)$ if there exists a \textit{Planck scale faithful embedding} \cite{Bombelli:1987aa}: that is, an injective map $\phi: C \hookrightarrow M$ satisfying the following conditions. 
 \begin{itemize}
\item[(i)] (Planck-scale uniform): The number of causal set elements embedded in any sufficiently large, physically nice region of $M$
 is approximately equal to the the spacetime volume of the region in fundamental volume units $V_f = z^4 V_p$ (see Remark C of Comment 3 in Section \ref{subsec:2asspn}). `Physically nice' means that the region contains large, approximately flat Alexandrov intervals and it has no Planck scale features such as very wiggly boundaries. 
\item[(ii)] (Order-preserving): Elements $x$ and $y$ of $C$ are ordered, $x\preceq y $, if and only  $\phi(x) \in J^-( \phi(y))$.
\item[(iii)] The characteristic distance over which the continuous geometry $(M,g)$ varies appreciably is everywhere much greater than the Planck length/time. (As mentioned in Comment 2 in Section \ref{subsec:2asspn}, this is a condition on any $(M,g)$ recovered by a Planck scale discrete theory X. We state it here as a reminder, and because it is in the original definition of faithful embedding for causal sets \cite{Bombelli:1987aa}.)
\end{itemize}
\end{mdframed}

\medskip

The DCC-C is the statement of the claim that when a faithful embedding of $(C, \prec)$ into $(M,g)$ exists, all the geometric information of $(M,g)$ on scales large compared to the Planck scale is encoded in $(C, \prec)$. 
Clearly, the concept of faithful embedding derives directly from Section \ref{subsec:causalcentral}'s Kronheimer-Penrose-Hawking-Malament theorem. If the macroscopic causal order of $(M,g)$ is recovered  from the microscopic order  of $(C,\prec)$ and the volume 
measure of the continuum, i.e. $(M,g)$'s one remaining `degree of freedom',  is recovered from the counting measure on $C$, then, the theorem suggests, that is sufficient to recover the geometry of $(M,g)$. 

The faithful embedding is the all-important concept of the discrete-continuum correspondence for causal sets (DCC-C): without it, the claim that causal sets `do the job' cannot be assessed. We emphasise that this correspondence via faithful embedding means that the causal set is not merely discrete but clearly, meaningfully, discrete \textit{at the Planck scale}. This condition on DPD---that the discreteness has a physical scale, which scale is the Planck scale---is meaningless at the level of the DPD \textit{in itself}. It only acquires meaning in the context of the discrete-continuum correspondence, as emphasised in Remark C of Comment 3 in Section \ref{subsec:2asspn}.  

Note also that this DCC-C  implies there is a whole class of causal sets \textit{ each} of which recovers $(M,g)$---just as there are many microscopically distinct molecular states of a gas each of which recovers the same continuum fluid state. 

In Comment 3 in Section \ref{subsec:2asspn} (Remark B), we noted that,  in order for theory X to recover GR, the set of DPD must recover an \textit{essentially unique} large-scale continuum spacetime. In the context of causal sets, we require that if $(C,\prec)$ faithfully embeds in two GR spacetimes $(M_1,g_1)$ and $(M_2, g_2)$ then $(M_1, g_1)$ and $(M_2,g_2)$ must be approximately isometric in the following way. Let $\phi_1: C \hookrightarrow M_1$
and $\phi_2: C \hookrightarrow M_2$ be the two faithful embeddings. Then there exists a $C$-preserving diffeomorphism $f: M_1\mapsto M_2$ that is an approximate isometry, where 
$C$-preserving means that $f\circ \phi_1 = \phi_2$  \cite{Bombelli:1987aa}.  
This is the so-called {\em Hauptvermutung} (``main conjecture'') of causal set theory. It says, essentially,  that the DCC-C as stated above \textit{works}: all the topological, differentiable and metrical structure of $(M,g)$ at scales large compared to the Planck scale is indeed encoded in any $(C, \prec)$ that is 
 faithfully embeddable in $(M,g)$. To state it another way, if the Hauptvermuting fails then so does the DCC-C. 
 
 The Hauptvermutung has been explored since the inception of causal set theory, often using the idea of {\em Poisson sprinkling}. Accordingly, we will now (i) introduce Poisson sprinkling (Section \ref{subsec:poisson}), and then (ii) describe some of the evidence for the Hauptvermutung (Section \ref{subsec:evidence}).
  
\subsection{The Hauptvermutung and Poisson sprinkling}\label{subsec:poisson} 

Much of the evidence for the Hauptvermutung---some of which is listed in Section \ref{extractgeomy} below---uses the concept of \textit{Poisson sprinkling} \cite{Bombelli:1987aa, meyer:1988, BOMBELLI1989226, Sorkin:1990bh,Sorkin:1990bj,Sorkin:2003bx}. A Poisson sprinkling is a process of selecting points of a given GR spacetime (Lorentzian manifold) $(M,g)$ that satisfies (iii) of DCC-C (Section \ref{subsec:kin}), at some density, and endowing the selected points with the order induced by the spacetime causal order, so as to produce a random causal set. If the density is unity in fundamental units, then typical Poisson-sprinkled causal sets faithfully embed in $(M,g)$ up to fluctuations in the number-volume relation that can be analysed. For example, in a Poisson sprinkling at Planckian density into a region of Minkowski spacetime the size of the observable universe, 
 the probability that there is an Alexandrov interval of radius equal to the size of an atomic nucleus that is empty of sprinkled points (in 
 which case it would violate (i) of the DCC-C) is approximately $10^{252} \times e^{-10^{72}}$
 \cite{Dowker:2003hb}.  So, according to the DCC-C, typical random Poisson-sprinkled causal sets are well approximated by $(M,g)$, at least when $(M,g)$ is of the size of the observable universe or smaller.  One corollary of this is that there is a huge number of causal sets that all recover the same GR spacetime. 

How special is the Poisson process? Can we say that a causal set $C$ faithfully embeds in GR spacetime 
$(M,g)$ \textit{only if} $C$ is a typical outcome of a Poisson sprinkling? Saravani and Aslanbeigi (SA)
studied this question and  proved that if the expected 
number of points chosen from each Alexandrov  interval in a point process in $(M,g)$ equals the volume
of the interval, then it cannot have smaller variance than the Poisson process (Theorem 1 of \cite{Saravani:2014gza}). So, causal sets that are typical outcomes of Poisson sprinkling are best at realising the number-volume correspondence for all Alexandrov intervals. SA point out that this 
is a stronger condition than needed for faithful embedding because we are only interested in the 
number-volume correspondence for regions large compared to the Planck scale. So SA  also provide evidence for the conjecture that in 2+1 and higher dimensions, Poisson sprinkling still provides the best number-volume correspondence if one demands only that the process have the correct mean for intervals larger than a certain volume.\footnote{More precisely: Saravani and Aslanbeigi explain why a known counterexample in 1+1 dimensions---a family of lattices in 1+1 dimensional Minkowski spacetime which provide, for large volumes, a better number-volume correspondence than Poisson sprinkling---should not have analogues in higher dimensions than 1+1 (Sections 3 and 4 of \cite{Saravani:2014gza}).}  
If this conjecture about Poisson sprinklings holds, it means that one can restate the discrete-continuum correspondence for causal sets, i.e. our DCC-C, as:  ``Causal set $(C, \prec)$ recovers GR spacetime $(M,g)$ at large scales if $(C, \prec)$ is a typical outcome of Poisson sprinkling at Planckian density into $(M,g)$''.

We will not rely on this conjecture about Poisson sprinklings, though we think it is likely to be true.\footnote{The DCC-C is often stated in the causal set literature in its Poisson sprinklings form.} But we will use the fact that typical Poisson sprinkled causal sets are faithful embeddings into $(M,g)$, and so they provide a huge class of causal sets on which to test the Hauptvermutung and the validity of the discrete-continuum correspondence, DCC-C.  

 Poisson sprinkling has been at the heart of progress on the concept of scale-dependent approximate isometry between Lorentzian manifolds.  Bombelli  \cite{Bombelli_2000} uses Poisson sprinkling to define a scale-dependent distance function on the space of finite-volume, distinguishing Lorentzian geometries; as follows.  He associates to each finite volume, distinguishing Lorentzian geometry, $(M,g)$,  the probability distribution over finite posets given by the Poisson sprinkling process into that geometry at unit density. He then proposes the Bhattacharyya angle, or statistical angle, between these probability distributions as a candidate, scale-dependent distance between geometries.\footnote{It remains a conjecture that the Bombelli function is a true distance in the sense that if the distance between two Lorentzian manifolds is zero then the manifolds are isometric, however this is highly plausible especially in the case of compact manifolds.} The Poisson distribution gives greatest weight to the posets whose cardinality equals $N \pm \sqrt{N}$, where $N$ equals the volume of the Lorentzian geometry, and very small weight to posets that are much larger than this. The Bombelli distance 
 function is therefore insensitive to structure on scales below that set by the density. Choosing the sprinkling density in the definition to be 10 rather than unity, say, increases the scale below which the distance function is insensitive to structure. 
 
 As well as its obvious potential usefulness for formulating a precise version of the Hauptvermutung for causal sets, the Bombelli distance function may be useful in any context where one wants to say that Lorentzian geometries are close above some scale.  It could provide a \textit{covariant} concept of a coarse-graining of a Lorentzian geometry, allowing spacetimes with different topologies and dimensions to be understood as close, above some scale: which is necessary to understand, for example, Kaluza-Klein theory more precisely. 

\subsection{Evidence for the Hauptvermutung}\label{subsec:evidence}
We now present a representative sample of the evidence for the Hauptvermutung, from the literature on causal sets. There are two broad kinds of evidence: (i) evidence from general theorems  (Section \ref{genev}); and (ii) evidence based on extracting geometrical information from a given causal set  (Section \ref{extractgeomy}). 

\subsubsection{General Evidence}\label{genev}

First we have, of course, the Kronheimer-Penrose-Hawking-Malament (KPHM) theorem,  the ``Ur Theorem'' of causal set theory. One can think of causal sets as the logical outcome of the heuristic that causal order is a more primitive organising principle even than space and time. Since, the KPHM theorem tells us that in the continuum, causal order is not---quite---sufficient for physics. But,  if the order is discrete, the missing scale information is included for free and measured by counting. Thus Sorkin, the foremost champion of the causal set approach to quantum gravity,  acknowledges the influence of Malament in coming to the understanding of just how much geometric information is encoded in the causal order of Lorentzian geometry \cite{Sorkin:1989re} and the influence of Riemann for mooting that a discrete manifold can contain its own metrical information (the relevant passages of Riemann's inaugural lecture \cite{Riemann:1868} are translated by Sorkin on pages 3 and 4 of \cite{Sorkin:2003bx}). Sorkin writes: ``[W]hat is especially appealing about causal sets is that their discreteness is essential to their ability to reproduce macroscopic geometry. If an infinite number of elements were present locally then the correspondence $V = N$ would lose its meaning and without it we could at best hope to recover the conformal metric, but not the volume-element needed to get from the latter to the full metric $g_{ab}$'' \cite{Sorkin:1990bj}.

Second, there are two theorems to the effect that Poisson-sprinkled causal sets are  Lorentz invariant \cite{Bombelli:2006nm,Dowker:2020aa}, thus supporting the intuition we have gained from fluid mechanics in which the microscopic molecular data do not break Euclidean invariance because this data is random---in direct contrast to a crystal whose regular arrangement of atoms does break  Euclidean symmetries.\footnote{Note that the concept of Lorentz invariance does not make sense for a set of DPD such as a causal set \textit{in itself}, not least because the DPD are discrete and Lorentz symmetry is a continuous symmetry. The claim that a faithfully embeddable causal set is Lorentz invariant can only be assessed \textit{in the context of} the discrete-continuum correspondence DCC-C.}  Taken together, the two theorems---more accurately, two theorem-types---constitute an almost complete formal proof that a typical causal set Poisson-sprinkled into Minkowski spacetime respects all of its Poincar\'e symmetry. The first theorem-type shows that, with probability one, a sprinkled causal set does not pick out a distinguished timelike vector in the approximating continuum, so it cannot prefer any inertial frame\cite{Bombelli:2006nm}. The second theorem-type shows that with probability one a sprinkled causal set does not pick out a distinguished lattice nor any other geometric structure in the approximating continuum whose symmetry group contains a translation \cite{Dowker:2020aa}.\footnote{Full Poincar\'e symmetry is respected by a sprinkled causal set  if the following conjecture is true: If $G$ is the Poincar\'e group and  $H$ is a subgroup of $G$ and $H$ does not contain a translation,  then the coset space $G/H$ has infinite volume (Section 4.3 of \cite{Dowker:2020aa}).} 

Third, although the Hauptvermutung is---crucially---about a continuum \textit{approximation} and \textit{not} a continuum \textit{limit}: if a continuum limit exists, it is evidence there is a continuum approximation nearby. We therefore count as evidence for the Hauptvermutung the result that a causal set that is Poisson-sprinkled into a distinguishing Lorentzian manifold of dimension $d>2$ recovers, in the limit of infinite density, the full continuum geometry with probability one \cite{Sorkin:1990bj, BOMBELLI1989226}. 

Finally we cite as general evidence the \textit{robustness} of causal sets, the  \textit{kinship} between a causal set and its approximating Lorentzian geometry and the \textit{explanatory power} of the DCC-C. By `robustness', we mean that there are no arbitrary elements
in the definition of a causal set. That is: the definition cannot be adjusted without utterly changing the character of a causal set. By `kinship', we mean that the causal set and approximating GR spacetime speak the same language. The 
 DPD of the causal set are \textit{physical} data in a rather straightforward way: they translate more or less directly to information in the continuum approximation---causal order and physical scale---that is physically meaningful in GR. Another way in which a causal set and its approximating manifold are like each other is their nonlocality: in a faithfully embedded causal set  the nearest neighbours  of any element $x$\footnote{A nearest neighbour to an element, $x$, in a causal set is an element $y$ that is related to $x$ and such that there is no element in the 
 order between $x$ and $y$. In the causal set literature such a pair $(x,y)$ is called a link.} are distributed in a neighbourhood of---and inside---the past and 
 future lightcone of $x$ in the approximating continuum \cite{Moore:1988zz}. Cf. Figure 1 in Section \ref{subsec:causalcentral}: the nearest neighbours to a causal set element embedded at the origin $p$ lie throughout  the striped region.  Indeed, in Minkowski spacetime, there are infinitely many nearest neighbours of any given element of a typical Poisson sprinkled causal set. Thus, as was stated in \cite{Bombelli:1988qh}: `Points in a Lorentzian manifold have metric neighborhoods which converge to the light cone rather than to the point itself. Causal sets simply reflect this characteristic of Lorentzian geometry.'  By `explanatory power', we mean that the DPD being a causal set would be an \textit{explanation} of the Lorentzian nature of spacetime in GR, since only a Lorentzian metric signature---of all the possible metric signatures---provides distinct past and future directions and hence a partial order.\footnote{With a Lorentzian $(- + + \dots +)$  signature the null vectors in the tangent space form two distinct cones---past and future pointing---and all timelike vectors are also either past or future pointing respectively depending on which cone they lie inside. With all plusses, $(++\dots ++)$,  all vectors in the tangent space are spacelike and there are no null vectors, no lightcones, at all.  In the remaining cases, the set of timelike (i.e. negative norm squared) vectors form a connected set, and so do not divide into two distinct classes and there is no way to define past and future pointing.} 
 
\subsubsection{Extracting geometry from a causal set}\label{extractgeomy}
We come now to evidence of a more specific kind: a representative set of examples of the explicit recovery of continuum spacetime geometric information from faithfully embeddable causal sets.
The idea is that, starting with a  Lorentzian manifold $(M,g)$, a faithfully embedded causal set $(C, \prec)$ is produced by Poisson sprinkling; and then information about the topology and geometry of $(M,g)$ can be extracted \textit{ from } $(C,\prec)$ \textit{alone}. 

 This accumulating body of evidence is a work in progress. In a typical example, a particular function, 
 $G$, of a causal set---and maybe e.g. one or more elements of the causal set or a partition of the causal set---is proposed as the underlying causal set quantity that recovers a particular continuum geometric quantity, $\mathcal{G}$,  in the approximating Lorentzian manifold $(M,g)$.  That causal set function is then evaluated on random faithfully embedded causal sets Poisson-sprinkled into a Lorentzian manifold $(M,g)$ at density $\rho$. This turns the causal set function $G$ into a 
 random variable $\mathbf{G}$. As a first step, (i), the expected value, $<\mathbf{{G}}>$ of the random variable and its limit as  $\rho \rightarrow \infty$ are calculated and shown to equal the continuum value of $\mathcal{G}$ in  $(M,g)$. Then, (ii) the value of $<\mathbf{G}>$ at finite $\rho$ is shown to be close to its limiting value when the discreteness scale set by $\rho$ is small compared to any curvature scale and (iii) the fluctuations around the expected value are shown to be small so that the continuum value of $\mathcal{G}$ can be read off from $\mathbf{G}$ evaluated on a \textit{single} faithfully embedded Poisson-sprinkled causal set $C$.  In some examples the fluctuation analysis remains to be done and what we have are promising first steps (i) and (ii). 
  
More examples and details are found in the reviews of causal set theory \cite{Sorkin:1990bh,Sorkin:1990bj,Sorkin:2003bx,Surya_2019}. 

\begin{itemize}

\item[] \textit{Causal past/future of points, Alexandrov intervals, spacetime volume of regions}:--- These continuum geometrical structures and volume information are more-or-less immediate from the discrete-continuum correspondence. For example, an Alexandrov interval in the continuum corresponds to a---large enough---\textit{order interval} of $C$, where an order interval in $C$ between $x,y\in C$ with $x\prec y$ is defined by $[x,y] := \{z \in C\,|\, x\prec z\prec y\}$. The spacetime volume of a region is given, up to Poisson fluctuations,  by the number of causal set elements comprising that region.

\item[] \textit{Spacetime dimension} There are a number of dimension discriminators/estimators. One is \textit{flat conformal dimension} which is based on a collection of causal sets, $C_d$, one for each dimension $d$ such that $C_d$ embeds in Minkowski spacetime of dimension $d$ but not $d-1$ (pp 8-9 of \cite{Sorkin:1990bh}, \cite{meyer:1988, Brightwell1989}). Another is \textit{midpoint scaling dimension} (page 85 of \cite{bombelli_thesis_1987},  \cite{Reid:2002sj}). 
 The Myrheim-Meyer (MM) dimension (page 12 of \cite{Myrheim:1978} and p. 52-53 of \cite{meyer:1988})  is a family of fractal, statistical dimensions.  `Fractal' here means that the dimension is not an integer in general, but will be close to an integer when the sprinkled causal set is large enough. The simplest MM dimension is the \textit{ordering fraction} of the causal set which is the fraction of pairs of elements that are related and for a causal set sprinkled into an Alexandrov interval of Minkowski spacetime the expected ordering fraction is a known monotonically decreasing function of dimension. A calculation of the variance shows that the ordering fraction accurately determines the dimension $d$ when the cardinality of the causal set is larger than $(27/16)^d$, making this dimension estimator very efficient \cite{Sorkinvideoinvitation3, Sorkinvideoinvitation4}. 
 
\item[] \textit{Timelike geodesics} A longest chain (i.e. a linearly ordered set of maximum cardinality) between two elements in a sprinkled causal set is a natural analog of a timelike geodesic in the continuum, as suggested by Myrheim (p. 6 of \cite{Myrheim:1978}). For flat sprinklings, the cardinality of a longest chain multiplied by the fundamental length, $L$---given by $\rho= L^{-d}$ where $\rho$ is the sprinkling density in dimension $d$---and by a dimension-dependent constant, $m_d$, of order 2 is close to the associated timelike geodesic's continuum-length (i.e. the proper time along it) for long enough geodesics (see pp. 6-7 in \cite{brightwell2015mathematics}, \cite{Brightwell:1990ha, Bollobas:1991}). 
Bachmat has shown that this result about longest chains extends in the continuum limit, 
$\rho\rightarrow \infty$,  for sprinklings into curved spacetime (Theorem 1.1 in \cite{Bachmat:2008}, Chapter 3.6.2 in \cite{Bachmat:2014}). It remains to be determined under what circumstances the longest chain is an accurate geodesic length estimator at \textit{finite} sprinkling density in curved spacetime. 

\item[]\textit{Approximately flat Alexandrov intervals} 

The MM dimension is an efficient dimension estimator for sprinklings into flat Alexandrov intervals and it should also be applicable to sprinklings into curved spacetime because every Lorentzian manifold `is approximately flat, locally'. 
This heuristic suggests a strategy, sketched by Bombelli on p. 83 of \cite{bombelli_thesis_1987}, for determining the dimension: namely, taking a sample of `mesoscale' order intervals in the causal set which would correspond to approximately flat Alexandrov intervals with volumes large compared to the Planck volume and small compared to any curvature scale, and calculating the 
MM dimension of such intervals. A stable value for the MM dimension from such mesoscale intervals uniformly covering the causal set would then be a good measure of the dimension of $(M,g)$. 

There is a subtlety here, however: even in a GR spacetime that satisfies the condition that the curvature scale is large compared to the Planck scale, not all Alexandrov intervals of the same mesoscale need be approximately flat. The reason is that when the timelike geodesic between the two endpoints of the interval is close to null, the \textit{volume} of the interval can remain small in magnitude compared to any curvature scale, even though the interval itself can stretch across a region of the spacetime in which the curvature varies significantly (see Figure 1 in \cite{Glaser:2013pca}). Such a `long and skinny' Alexandrov interval may not be approximately flat even though it has small volume due to the presence of nonzero Weyl tensor; and the MM dimension of the corresponding order interval in a sprinkled causal set would be inaccurate. However, in the continuum, for every point $p$ in a GR spacetime, there is a large family of mesoscale Alexandrov intervals each containing $p$, which are all approximately flat and are related to each other by approximate Lorentz boost symmetries. In order for a causal set $C$ to be manifold-like, therefore, for each element $x$ there must be a corresponding set of mesoscale order intervals in $C$ that contain $x$ and are approximately flat.  In principle, one could search through all mesoscale order intervals containing $x$ and check that some large subset of them give
a stable result for the MM dimension. If one can cover the whole causal set with such
mesoscale order intervals with stable MM dimension, that is evidence of manifold-like-ness.

 In the sprinkled causal set identifying the order intervals that correspond to approximately flat Alexandrov intervals, is clearly important. One way to do this is by comparing the abundances of small sub-order intervals within each candidate mesoscale order interval and comparing it to what they would be in sprinklings into flat intervals of dimension $d$ \cite{Glaser:2013pca}. As the abundances depend on dimension, this would in itself be, effectively, a dimension estimator. 

\item[] \textit{Spatial topology and geometry} The Lorentz invariance and consequent non-locality of causal sets mean that they struggle to reproduce spacelike geodesics and spacelike geodesic distance, as explained in \cite{Brightwell:1990ha}.  Some progress has been made, however, with promising evidence for one proposal for flat sprinklings \cite{Rideout_2009}.
 
Where spatial information pertains to a Cauchy surface, however, this anchors the problem and tames the nonlocality somewhat, so that 
more can be done. For example, a Cauchy surface can be associated to a \textit{thickened maximal antichain}, where an {\em antichain} is a set of mutually unordered elements and the thickening includes all elements, $y$ to the future, say, of the antichain which have no more than $k$, say, elements between $y$ and the antichain.  Information about the spatial topology \cite{Major:2006hv, Major:2009cw}
and geometry  \cite{Eichhorn_2019} of the associated Cauchy surface can be deduced from the order restricted to the thickened antichain.

\item[] \textit{Ricci scalar curvature and Ricci tensor components} 
The discovery of a linear operator on 
scalar fields on causal sets that recovers the scalar D'Alembertian on Minkowski spacetime was a major breakthrough  \cite{Sorkin:2007qi}. In fact, a one parameter family of such operators exists (eq. 7 of   \cite{Sorkin:2007qi}) with the parameter being a non-locality or smearing scale 
that acts to tame the fluctuations around the expected value.  It was then found that the mean of this linear operator acting on a constant 
field equal to $-2$  on a causal set sprinkled into an approximately flat interval of curved spacetime is close to the Ricci scalar curvature \cite{Benincasa:2010ac, Belenchia:2015hca, Dowker:2013vba}. A systematic analysis of the fluctuations of this scalar curvature estimator remains to be done. 

There are a number of ideas for the recovery of more components of the curvature tensor. 
For example, Myrheim showed that the deviation of the continuum volume of an approximately flat
Alexandrov  interval from the Minkowski spacetime value  depends on both the Ricci scalar and the component of the 
Ricci tensor in the direction of the timelike geodesic between the ends of the interval \cite{Myrheim:1978}.  
Once one has recovered the Ricci scalar, therefore, this will allow the recovery of the component $R_{tt}$ along a particular timelike geodesic. 

\item[] \textit{Spacelike hypersurface volume and extrinsic curvature} Another causal set concept that can correspond to a hypersurface in the continuum is a \textit{partition} of the causal set into two parts (i.e. two jointly exhaustive and mutually exclusive subsets). For a spacelike hypersurface, information about the geometry of that hypersurface---the 3-volume and the integral of the extrinsic curvature---can be recovered from certain causal set functions (eqs (9) and  (12) of \cite{Buck:2015oaa}): at least in four dimensions, where there is evidence that the fluctuations are small (Section 3.4 of  \cite{Buck:2015oaa}). 

\item[] \textit{Area of causal horizon} In the continuum, a causal horizon---a generalisation of the concept of black hole event horizon that includes cosmological horizons and acceleration horizons---is the boundary of the causal past of a future infinite timelike worldline. A timelike worldline corresponds to a maximal 
chain. So in the causal set, the causal horizon corresponds to the partition of the causal set into the past of the chain and the complement of the past of the chain. Where there is a causal horizon intersected by a spacelike hypersurface in the continuum, this corresponds to two partitions of the causal set: intersecting these partitions gives a partition of the causal set into four parts. 
The continuum area of the intersection of the causal horizon with the spacelike hypersurface can be shown to be equal to the continuum limit of the expected value of a causal set function depending on this partition of the causal set into four parts (Eq (2.1) of  \cite{Barton:2019okw}). An analysis of the fluctuations for individual causal sets remains to be done. We can hope that, since the geometric quantity recovered is an area and therefore itself an integral of a local quantity,  like the 3-volume above, the fluctuations will be well behaved, at least in four dimensions. 
 
\end{itemize}

\section{Alternatives to causal sets?}\label{sec:lattices}

In this Section,  we turn to supporting Claim 2 from Section \ref{subsec:2claim}: that causal sets are the only proposal, known so far, for sets of fundamentally discrete Planck scale DPD that recover GR spacetimes. 

We make the claim---despite our of course not knowing all the relevant papers---on the basis of an argument that the most familiar discrete manifolds, namely lattices and simplicial complexes, which might seem to be counterexamples to our Claim, do {\em not} do the job. 

We present this argument in Sections \ref{sec:lorlatt} and \ref{subsec:void}. In short, we will argue that (a natural formulation of) the Hauptvermutung---i.e. a natural discrete-continuum correspondence---for a Lorentzian simplicial complex {\em fails}. Since it is the \textit{Lorentzian} character of the desired recovered continuum that is their downfall, we also briefly discuss (in Section \ref{subsec:eucl}) why this objection to lattices and simplicial complexes as DPD does not apply to the Riemannian case. 

Before giving the argument, we first clarify what we mean by simplicial complex. As already mentioned in Remark C of Comment 3 in Section \ref{subsec:2asspn} and Remark B in Section \ref{subsec:2claim}: there are two conceptions of a simplicial complex, which we call the \textit{geometric simplicial complex} and the \textit{combinatorial simplicial complex}. The former is (in dimension $d$) a union of $k$-simplices, for $k= 0,1,2,\dots d$, which are pieces of $k$-dimensional flat space (either Euclidean space if this is a Riemannian geometric simplicial complex, or Minkowski spacetime if this is a Lorentzian geometric simplicial complex) satisfying certain conditions. If this geometric simplicial complex satisfies certain further conditions it will be a piecewise flat (PF) $d$-dimensional manifold. On the other hand, a combinatorial simplicial complex is discrete: it is a collection of vertices, pairs-of-vertices, triples-of-vertices \textit{etc.} satisfying a set of conditions. \textit{Geometrical} simplicial complexes do not satisfy our requirement of being discrete data sets.

Since this is an important point, we give a further argument. Suppose, for the sake of a {\em reductio}, we allow  geometrical simplicial complexes as DPD. Consider the particular case in which the Lorentzian geometry to be recovered is 4-dimensional Minkowski spacetime itself, and the DPD is a {\em geometric} simplicial complex that is some particular triangulation of Minkowski spacetime. The DPD in this case, then, simply \textit{is} Minkowski spacetime. But Minkowski spacetime is {\em not} discrete, let alone discrete at the Planck scale. More generally: a piece of  a continuum flat spacetime is not the absence of data: it is substantial, it is a continuum with all its topological, differentiable, metrical and causal structure. It contains physical data of the sort we are excluding with our Assumption 2 (Section \ref{subsec:2asspn}). However, a combinatorial simplicial complex---with or without edge lengths, triangle areas, or other decorations---\textit{is} discrete. So when we refer to a simplicial complex, we will mean a combinatorial simplicial complex (unless specified otherwise). 

Thus we emphasise again (cf. Remark B in Section \ref{subsec:2claim}) that our argument is no obstacle to a {\em geometric} Lorentzian simplicial complex being considered, and used to approximate a continuum Lorentzian geometry.\footnote{Note the direction of the approximation relation here.} 
Indeed the triangulation of Minkowski spacetime just considered  is simply a case in which the approximation of the continuum spacetime is \textit{perfect}.

\subsection{The combinatorial Lorentzian simplicial complex does not do the job}\label{sec:lorlatt}

Consider a 4-dimensional Euclidean geometric simplicial complex,  
 $\bar{S}$ that is also a piecewise flat Riemannian manifold. 
 Let $S$ be the  \textit{combinatorial} simplicial complex corresponding  to $\bar{S}$.   Let each edge (1-simplex) of $S$ be decorated by a \textit{Lorentzian edge-length}, where a Lorentzian edge-length is a timelike, null (zero) or spacelike length and also a future pointing direction in the case of timelike or null edges. Denote these decorations collectively by $dec$.  
We call $(S,dec)$ a {\em Combinatorial Lorentzian Regge Complex} (CLRC).\footnote{We call this a Regge Complex because in Regge calculus, the edges of a geometric simplicial complex are decorated
by edge-lengths and those edge-lengths determine the geometry of each simplex of the complex if the interior metric is flat, as is assumed in Regge calculus \cite{Regge:1961, sorkin_thesis}. In Regge calculus there are consistency conditions on the Lorentzian edge-lengths of a simplex if the interior flat metric is to have a Lorentzian signature (\cite{sorkin_thesis}). We do not impose signature conditions on the Lorentzian edge-lengths, $dec$, at this stage of defining a CLRC. Signature conditions will be (implicitly) imposed within the upcoming proposal for a discrete-continuum correspondence between a CLRC and a GR spacetime.}

Now let the {\em discrete-continuum correspondence}---which we will call: {\bf DCC-CLRC}---be as follows. (Note the contrast with Section \ref{subsec:kin}'s discrete-continuum correspondence for causal sets,  DCC-C.)

\medskip

\begin{mdframed}
\noindent A CLRC, $(S, dec)$,  recovers the GR Spacetime $(M,g)$ if the following conditions hold.
  \begin{itemize} 
 \item[(i)] There is a piecewise flat manifold $\bar{S}$ that is a triangulation of 
 $M$ such that $S$ is the combinatorial simplicial complex corresponding to $\bar{S}$.
\item[(ii)] There is a homeomorphism $f: \bar{S} \rightarrow M$  such that the $f$-image of each edge of $\bar{S}$  is a---spacelike,  timelike or null---geodesic in $(M,g)$. $f$ induces an embedding of the vertices of $S$ in $(M,g)$. 
\item[(iii)] The spacetime length and future pointing direction (if appropriate) of the $f$-image of each edge of $\bar{S}$ in $(M,g)$ equals the Lorentzian edge-length of the corresponding edge of $(S, dec)$ up to some tolerance.
\item[(iv)] The edge-lengths are no greater that a few Planck units. 
\item[(v)] The characteristic distance over which the continuous geometry $(M,g)$ varies appreciably is everywhere much greater than the Planck length/time. 
\end{itemize}
\end{mdframed}

\medskip

We admit that this definition, DCC-CLRC, is not the only possible definition of a Planck scale discrete-continuum correspondence for CLRCs. But it is reasonable and it builds on one's intuitions  about geodesic domes and similar examples, as discussed in Section \ref{subsubsec:analogies}.  It is also intuitive in the sense that the metric in GR is, more often than not, given by a line element, $ds^2 = g_{\mu \nu}(x)  dx^\mu dx^\nu$ and the geodesic edge-length data corresponds directly to continuum metric information on short line segments. 

We now show that the Hauptvermutung for DCC-CLRC {\em fails} by providing a counterexample. More specifically, we will exhibit a CLRC that, according to DCC-CLRC, ``recovers'' {\em both}  Minkowski spacetime {\em and} a spacetime with a gravitational wave burst (and, in fact, many spacetimes with different gravitational wave bursts). In view  of our requirement of essential uniqueness (Remark B in Comment 3 of Section \ref{subsec:2asspn}), this ``double win'' is a failure.  

\bigskip

Let $\bar{S}$ be the triangulation 
of 4-d Euclidean space, $\mathbb{R}^4$, used in \cite{ROCEK198131}.\footnote{Note that Rocek and Williams use the word `edge' to mean edges of the underlying lattice only and use ``face diagonals'', ``body diagonals'' and ``hyperbody diagonal''
to refer to the other 1-simplices. But we use  `edge' to refer to any 1-simplex. Note also that Rocek and Williams 
are working in Riemannian signature, so that the edge-lengths they quote are Euclidean edge-lengths.} 
The vertices of the triangulation are the vertices of the integer lattice. The unit 4-$d$ hypercube at the origin with the 16 vertices $\{(0,0,0,0), (0,0,0,1), (0,0,1,0), (0,0,1,1), \dots (1,1,1,1) \}$  is triangulated into
24 4-simplices. Each of these 24 4-simplices is the convex hull of a monotonic path on the lattice from $(0,0,0,0)$ to $(1,1,1,1)$. There are 15 edges  or ``lattice vectors''  that point from $(0,0,0,0)$ to each of the other vertices of the hypercube and these lattice vectors are labelled 1-15 by reading their vector components as binary numbers, 
so that the lattice vector $(1,0,0,0)$ is labelled 8, for example.
This fundamental triangulated unit hypercube is translated all over the integer lattice---one such 
hypercube at each lattice vertex---to give $\bar{S}$. We take the combinatorial complex, $S$, corresponding  to $\bar{S}$ and label its edges with 
Lorentzian length labels given in the following table in units of the Planck length, $l_{p}$,  and Planck time,
$t_{p}$: 
\vskip0.5cm

\begin{tabular}{|c|c|c|c|c|}
\hline
Timelike (1) & Null (0) & Spacelike (1) &Spacelike($\sqrt{2}$)&  Spacelike($\sqrt{3}$) \\
\hline
8: (1,0,0,0)& 9: (1,0,0,1)& 1: (0,0,0,1)&3: (0,0,1,1)&7: (0,1,1,1)\\
~& 10: (1,0,1,0)& 2: (0,0,1,0)& 5: (0,1,0,1)& ~\\
~& 12: (1,1,0,0)& 4: (0,1,0,0)& 6: (0,1,1,0)& ~\\
~& ~& 11: (1,1,0,1)&15: (1,1,1,1)& ~\\
~& ~&  13: (1,1,0,1)& ~& ~\\
~& ~& 14: (1,1,1,0)& ~& ~\\
\hline

\end{tabular}
\vskip0.5cm
  Also, the timelike and null edges are future-directed in the direction of increasing first coordinate.
 This, then, is our CLRC, $(S, dec)$. It corresponds to the geometric Lorentzian simplicial complex in 
 \cite{Williams:1986} which is the triangulation of Minkowski spacetime corresponding to 
$\bar{S}$ 
 and the Lorentzian edge-lengths $dec$ are the Minkowski spacetime proper lengths of the geodesics which are straight lines in the canonical inertial coordinate system defined by the fact that $M = \mathbb{R}^4$

\bigskip

According to the DCC-CLRC,  $(S, dec)$ recovers Minkowski spacetime, $(M_{mink}, g_{mink})$. This is straightforward but for completeness, let us explicitly check that the conditions (i) to (v)  of DCC-CLRC are satisfied. (i): By definition of $S$, the required $\bar{S}$ is the triangulation of $\mathbb{R}^4$ described above. (ii):  The manifold for Minkowski spacetime is $M_{mink} = \mathbb{R}^4$ and its canonical coordinate system is 
inertial. The triangulating map $f: \bar{S} \rightarrow M_{mink}$ is the identity from $\mathbb{R}^4$ to $\mathbb{R}^4$ and the edges of $S$ are embedded, via this map, as straight coordinate lines between vertices. (iii): These straight coordinate lines are geodesics in Minkowski spacetime and have the spacetime proper lengths matching the Lorentzian edge-lengths $dec$ given in the table. Finally,  (iv) and (v) are satisfied.  

Now, this simplicial complex $(S, dec)$ violates Lorentz invariance---just as its Riemannian version violates Euclidean symmetry and this may be enough reason for some to reject it as possible DPD in a quantum gravity theory. But we have a much more serious charge: $(S, dec)$ does not recover Minkowski spacetime at all because there are many other GR spacetimes that, according to the discrete-continuum correspondence, DCC-CLRC, are ``recovered'' by $(S, dec)$. 

To construct an example of such a spacetime, we will use the widely known
 fact that the distribution of the embedded vertices of $S$ induced by $f$
  is not actually uniform in Minkowski space. The distribution only \textit{seems} uniform in the frame preferred by the complex in 
which the images of the vertices under $f$ form the integer lattice. When boosted, in the $z$-direction say, this distribution is revealed to be very far from uniform: see for example Figure 1(a) of \cite{Saravani:2014gza}.
Indeed the boosted distribution of $f$-image-vertices has large \textit{voids}: large, physically nice, regions of spacetime in which there are no $f$-image-vertices. By `large, physically nice regions' we mean regions of spacetime that contain approximately flat Alexandrov intervals of  spacetime volume large on the Planck scale. Thus arises the possibility that there is curvature or physics---\textit{e.g.} a gravitational wave burst---in the voids, that does not register (get encoded) at all in the edge-lengths that decorate the CLRC. This will be the basis of our counterexample.

\begin{figure}[h!]\label{figure2} 
\centering
{\includegraphics[scale=0.5]{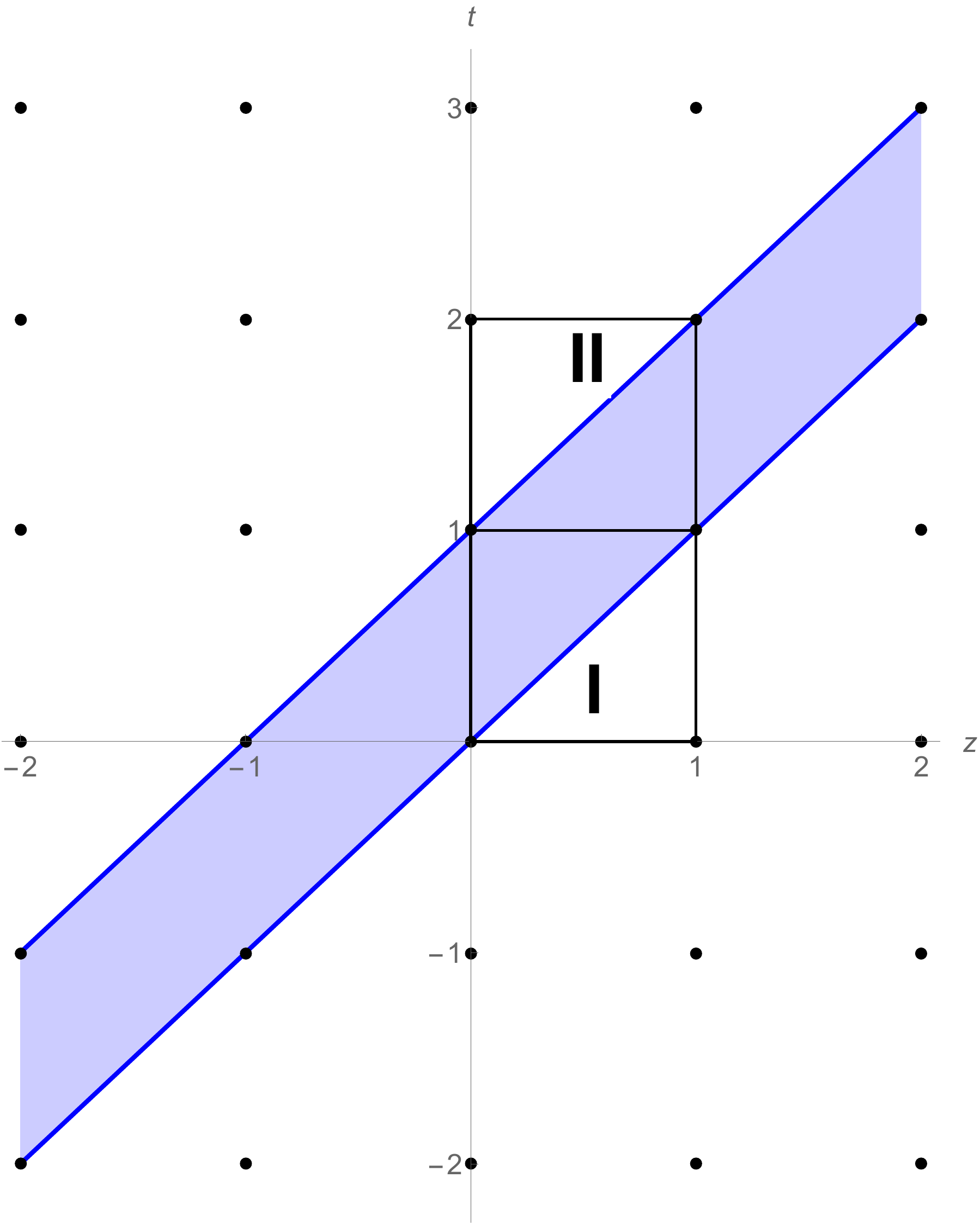}}
\caption{The diagram represents both Minkowski spacetime and the spacetime (\ref{eq:gwburst}) with a plane gravitational wave burst. 
The integer lattice of image-vertices in the plane $x=y =0$ is shown and most edges are not drawn. 
 The shaded region is $0 < t-z < 1$ and is empty of vertices. In the gravitational wave spacetime the shaded region is the support of the wave burst. The labels I and II refer to two hypercubes at $(0,0,0,0)$ and at $(1,0,0,0)$ respectively which intersect the support of the wave burst. All other coordinate-hypercubes that intersect the 
 support of the wave burst are isometric either to I or to II. 
}
\end{figure}

In Figure 2 is shown the $x=y=0$ plane of the integer lattice in Minkowski space---with inertial coordinates $\{t,x,y,z\}$---with the $x$ and $y$ dimensions suppressed. Let $u := \frac{1}{\sqrt{2}}(t-z)$ and $v :=  \frac{1}{\sqrt{2}}(t+z)$ 
be null coordinates in the $z$ direction.  The shaded region in Figure \ref{figure1} is the open region $ 0< u< \frac{1}{\sqrt{2}}$ which is crossed by image-edges of the complex but is empty of image-vertices: it is a void.  If the lattice is boosted in the $z$-direction, the distribution in the $u$ direction stretches by some factor $b$
 and by factor $b^{-1}$ in the $v$ direction. When the $\gamma$ factor of the boost is very large then 
 $b \approx 2\gamma$. 
 Choosing $\gamma = 10^{44}$, say, the height of the void in the 
 time coordinate becomes of the order of one second and contains
 Alexandrov intervals of height one second: the void is physically nice and large. 

We now exhibit a GR spacetime which is also `recovered' by $(S, dec)$ according to DCC-CLRC. Let $(M,g)$ be a perturbation of Minkowski spacetime that is a plane-fronted, transverse traceless gravitational wave burst in the $z$-direction:
\begin{align}\label{eq:gwburst}
ds^2 = & - dt^2 + dz^2   + ( 1 + \epsilon h(u)) dx^2 + (1- \epsilon h(u)) dy^2 \,,
\end{align} 
where the support of $h(u)$ is $ 0 < u < \frac{1}{\sqrt{2}}$,  and $u := \frac{1}{\sqrt{2}}(t-z)$ as before. The spacetime is flat for $u\le 0$ and $u \ge \frac{1}{\sqrt{2}}$.  The manifold $M$ is $\mathbb{R}^4$: the point $(t,x,y,z)\in \mathbb{R}^4$ is the point in $(M,g)$ with coordinates $(t, x, y, z)$, in the obvious way. 

We now check that the conditions (i) to (v) of DCC-CLRC hold.
(i): By definition of $S$, the required $\bar{S}$ is the triangulation of $\mathbb{R}^4$, exactly as in the case of Minkowski spacetime. (ii): Let  
$f: \bar{S} \rightarrow M$ be a homeomorphism that is a perturbation of the identity map chosen such that:
\begin{itemize}
\item[(a)] $f$ is the identity outside the support of $h(u)$ so that $f$ maps the vertices of $\bar{S}$ to the points with
integer coordinates in $(M, g)$ and the support of $h(u)$ is therefore empty of image-vertices. 
\item[(b)] $f$ maps the edges of $\bar{S}$ to the geodesics between the corresponding embedded vertices in $(M, g)$. 
\end{itemize}
Such a homeomorphism exists because (\ref{eq:gwburst}) is a perturbation of Minkowski spacetime
and there is a unique geodesic between two vertices of one coordinate-hypercube. 
(iii): Most of the geodesics are between vertices that lie outside the support of $h$ where the metric is flat and so they are straight lines in the coordinates  $\{t, x, y, z\}$ of (\ref{eq:gwburst}) and have the same length as in Minkowski spacetime. For the geodesic edges that cross the support of $h$, it can be shown that they have the same 
length as in Minkowski spacetime to first order in $\epsilon$, provided that we  choose $h$ to satisfy the constraint:
\begin{equation}\label{vanish1}
\int_0^\frac{1}{\sqrt{2}} du \, h(u) = 0\,.
\end{equation}
The details of the calculation are given in the Appendix \ref{appendix}. Furthermore, condititon (iv) holds. 
The only condition that might give one pause is (v). We deal with the doubt that (v) holds as a response to Objection 1, below. 

There are many choices of wave packet contour $h(u)$ that satisfy the criterion (\ref{vanish1}); and
for each choice, there is a GR spacetime that is ``recovered'' by $(S, dec)$ according to DCC-CLRC. Thus the requirement of essential uniqueness (Remark B of  Comment 3 in Section \ref{subsec:2asspn}) fails. And so we conclude that DCC-CLRC does not work as a statement of the recovery of a GR spacetime from a CLRC. 
 
 \bigskip

We now consider three possible objections to this conclusion. (Some other objections will be addressed in Section \ref{sec:loop}.) 
\\

\noindent \textit{Objection 1}:--- The spacetime (\ref{eq:gwburst}) does not satisfy condition (v) of
the DCC-CLRC because it varies on the scale of the discreteness; and so it should be dismissed out of hand.  \\
\noindent \textit{Reply}: Were this a Riemannian geometry, this objection would be valid. 
However, the gravitational wave spacetime is Lorentzian: and in a
boosted coordinate system, the support of the wave burst
is revealed to be a physically nice, large region of spacetime in which the contour of the burst varies slowly on the discreteness scale.

 For definiteness, consider the coordinate transformation, $\{t,x,y,z\}\rightarrow \{t', x, y, z'\}$ which is a boost in the $z$ direction such that $\{u, v\} \rightarrow \{u', v'\}$ where
$u' = b u$ and $v' = b^{-1} v$ where $b \approx 2 \gamma$ and the $\gamma$ factor is of order 
$10^{44}$. Then the metric  (\ref{eq:gwburst}) in the new coordinates is
\begin{align}\label{eq:gwburstprime}
ds^2 = & - d{t'}^2 + d{z'}^2   + ( 1 + \epsilon H(u')) dx^2 + (1- \epsilon H(u')) dy^2 \,,
\end{align} 
where $H(u') = h(u)$ and  the support of $H(u')$ is $ 0 < u' < \frac{1}{\sqrt{2}} b $. 
This is the same GR spacetime, expressed in different coordinates. The derivatives of $H$ with respect to $t'$ and $z'$ are small,  the curvature components are small on Planckian scales and the support is a physically nice region containing approximately flat Alexandrov intervals of proper height of 
order one second. The concrete definition given in Comment 2 of Section \ref{subsec:2asspn} is satisfied because $\{t',x, y, z'\}$ is approximately an inertial coordinate system everywhere in the support of the burst and in these coordinates the Riemann tensor is 
given approximately by second derivatives of  $\epsilon H(u')$.\\

\noindent \textit{Objection 2}:--- The spacetime (\ref{eq:gwburst}) is unphysical: it is only a vacuum solution at linear order; the wave has no physical scale being a free wave packet on a Minkowski spacetime background; and it is infinite in extent. It is not included in the collection of GR spacetimes that must be recovered by theory X, mentioned in Comment 2 in section \ref{subsec:2asspn}. A realistic gravitational wave burst would not be a counterexample to the CLRC Hauptvermutung. \\
\noindent \textit{Reply}: In a more realistic spacetime containing a gravitational wave burst, the finite spacetime support of the packet, the existence of other curvature structure in the universe,  and the discreteness of the---assumed---underlying physical data all place limits on the Lorentz transformations that are relevant physical symmetries. In a realistic spacetime, the $\gamma$ factor of a Lorentz transformation that is relevant here as a physical symmetry is bounded above by the ratio of the scale on which the curvature starts to be non-negligible to the scale of the discreteness. In the observable universe, for example, the largest possible $\gamma$ factor that needs to be considered is smaller than $10^{60}$, the Hubble horizon size in Planck units.\footnote{If one only needs to recover GR spacetimes of the size of the observable universe in which there is a limit to the $\gamma$ factors of the Lorentz transformations one need consider, a lattice or CLRC 
 with spacing equal to $10^{-60}$ times the Planck length/time might be able to provide all the geometric information to recover such a GR spacetime. However, such a lattice/CLRC would violate the assumption of Planck scale discreteness.}
 
 But the counterexample we have given  does not depend on Lorentz symmetry with arbitrarily high $\gamma$ factors. It depends on there being a symmetry with $\gamma$ factor of $10^{44}$---or lower if one considers wave bursts of, say, milliseconds or shorter duration to be part of GR. For the conclusion drawn from the counterexample to be valid, it is enough that there exist two realistic GR spacetimes---to be recovered by the theory X---that differ only by the presence or not of a plane gravitational wave burst in a region of spacetime that is: close to flat, large compared to $10^{44}$ times the Planck length, and such that a Lorentz transformation with 
gamma factor $10^{44}$ is an approximate symmetry. For example, the spacetime between galaxies with and without a millisecond gravitational wave burst would satisfy these conditions \cite{Abbott_2019}. 
\\

\noindent \textit{Objection 3}:--- The DPD could be a \textit{set} of combinatorial simplicial complexes, such that for each local Lorentz frame there is a member of the set that is uniform in that frame.\\
\noindent \textit{Reply}:  The discrete-continuum correspondence for such a set of data would involve an embedding of the vertices of all the simplicial complexes into the recovered spacetime. With one complex for each frame,  that would 
result in a density for the vertices' embedding much higher than Planckian, if not infinite. Such data
would contradict our assumption that the discreteness is Planckian in scale.

\subsection{Avoiding voids?}\label{subsec:void}

The failure of the Hauptvermutung means that the discrete-continuum correspondence  DCC-CLRC does not hold good: the continuum is \textit{not} recovered by a CLRC that satisfies the conditions of DCC-CLRC. The underlying reason for the failure of DCC-CLRC in the example above is the presence of a large, physically nice void region in the putative continuum, for which the only discrete data are
the lengths of the geodesic edges that cross the void; but that information is not nearly rich enough to encode the Lorentzian geometry in the void. (Recall the Reply to Objection 1 in Section \ref{sec:lorlatt}.) Moreover, there are voids \textit{everywhere}, not just the one highlighted by the gravitational wave burst example. We conclude: \textit{the CLRC, $(S, dec)$, of our counterexample is not approximated by a Lorentzian manifold at all.} 

One  might attempt to avoid this conclusion by eliminating voids: that is, by amending the discrete-continuum correspondence for CLRCs so as to require the number-volume correspondence  for the embedding of the vertices of the CLRC in the manifold,  i.e. condition (i) in DCC-C of Section \ref{subsec:kin} (Planck-scale uniformity). But there are two problems with this suggestion.

Firstly, if the vertices and the timelike and null edges of the simplicial complex form a directed acyclic graph then its transitive completion will  \textit{be} a causal set. Then, if Planck-scale uniformity holds for the embedded vertices of the complex, then either this causal set will be faithfully embedded
in the GR spacetime  or the order relation of the causal set defined by the directions on the complex's edges  will not be consistent with the spacetime causal order of the embedded vertices in the GR spacetime. The latter would be an inconsistency between the DPD and the continuum and would be hard to work around. The former case implies that for a CLRC to provide a genuine alternative set of DPD to a causal set whilst incorporating Planck-scale uniformity of embedding of vertices, the CLRC would have to have mostly or all spacelike edges. Even in that case, one could argue that from the information about the spacelike edge-lengths a unique causal order on the vertices of the CLRC should be deducible and so again with the imposition of Planck-scale uniformity on the embedded vertices of the complex, the complex would be effectively providing a faithfully embeddable causal set. 

A second, stronger argument against the possibility of eliminating voids by adding a requirement of Planck-scale uniformity is that a triangulation of a Lorentzian spacetime
cannot be constructed from a properly uniform distribution of embedded vertices, without introducing a frame. We note in particular that the
constructions of Voronoi graphs and Delaunay triangulations from a set of embedded vertices, which are so natural in Riemannian geometry (e.g. \cite{itzykson:1984}), do not work for Lorentzian manifolds,
and a supplementary frame must be introduced and used---as, for example, in \cite{tdlee:1985}. 
More generally, the basic obstruction to a Lorentz-invariant simplicial complex is the 
 contradiction between (i) the local nature of a simplicial complex with its concept, inherent in its structure, of nearest-neighbour vertices joined by 1-simplices and (ii) the non-locality of the Lorentzian geometry. 
For example, in a uniform, Poisson distribution of embedded vertices, a given vertex will have a nonlocal 
set of hugely, if not infinitely, many \textit{physically nearest neighbours},  hugging its 
past and future lightcones---as discussed in Section \ref{genev}---which a simplicial complex cannot accommodate. 

\bigskip

Finally, we note that the counterexample and all the above arguments apply also to Lorentzian  low valence graphs\footnote{Causal sets are graphs---transitive directed acyclic graphs---and the condition of low valency is to make sure these graphs are genuine alternative DPD sets to causal sets.}
 including the special cases of Lorentzian ``lattices" (graphs-with-symmetry). For consider: a CLRC contains within itself 
a Lorentzian low valence graph which is its 1-skeleton,  the set of its vertices and edges. So we can take the decorated 1-skeleton of the  CLRC $(C, dec)$ of our counterexample constructed from the integer lattice, and use it as a counterexample to any proposed DCC for Lorentzian low valence graphs.

\subsection{The contrast with the Riemannian case}\label{subsec:eucl}

We make no strong claim one way or the other about whether a Riemannian geometry can well approximate a Riemannian combinatorial simplicial complex or graph. However, we find plausible the claim that it can; and w will briefly give our reasoning here since it is illuminating to contrast with the Lorentzian case. 

Euclidean transformations of an embedding of vertices in Euclidean space preserve the uniformity---or non-uniformity---of the embedding in any Cartesian coordinate system. For example, consider the same combinatorial simplicial complex, $S$, as above, with its vertices on the integer lattice, and with its edges decorated with the Euclidean edge lengths. The distribution of embedded vertices is uniform in every Cartesian coordinate system: the number-volume correspondence holds for the number of vertices embedded in any sufficiently large, physically nice region of 4-dimensional Euclidean space. 
What about the ``void region'' identified above in the Minkowski space case: isn't it still there in the Euclidean case? That void region causes trouble in the Minkowski case because it is of large volume and physically nice. (Recall the Reply to Objection 1 in Section \ref{sec:lorlatt}.)  In Euclidean space, that void region has the same large volume but it is not physically nice:  it has physical, geometric structure on scales smaller than the discreteness scale. For its width, in the $z$-direction or the $t$-directions, is unchanged by any Euclidean translation or rotation.

As regards the recovery of topology and curvature, there are longstanding concepts of scale-dependent Hausdorff dimension and spectral dimension for combinatorial Riemannian simplicial complexes. There are also more recent proposals for associating curvature to a graph that give zero curvature for regular lattices uniformly embeddable in flat space \cite{Ollivier:2007, Ollivier:2009}; and there is evidence that the Ollivier curvature of a graph 
embedded in a Riemannian manifold tends to the continuum curvature when the density of the embedded vertices of the graph tends to infinity \cite{vanderhoorn2020ollivier}.

\section{Replies to objections}\label{sec:loop}
With our main argument now completed, we will in this Section reply to two objections that might be made. (The next Section will reply to another, for which our reply will need a brief discussion of the dynamics of causal sets.) But before stating the two objections, we stress for clarity that of course, one might reject one or more of our argument's assumptions. Here are two examples.

Thus someone might say: `One does not need to recover General Relativity'. To which, our reply is in two parts. First: This is a denial of Assumption 1, as discussed in Comments 1 and 2 of Section \ref{subsec:2asspn}. But fair enough: some kind of alternative theory of gravity (ATG) in which Lorentz invariance is violated, e.g. in which there is a preferred frame related to an aether field or a class of ``observers'',  {\em might} replace General Relativity as a better theory of gravity and of spacetime at macroscopic scales. And in that eventuality, our argument as it stands does not apply. But, second:  if it were claimed that such an ATG at macroscopic scales can be recovered from a Planck scale discrete theory of quantum gravity, then our argument provides a framework for investigating whether and how a Planck scale discrete theory of quantum gravity might recover an empirically adequate ATG at macroscopic scales. For example, in the case of an ATG whose kinematics includes a Lorentzian spacetime, $(M, g)$ and a preferred timelike vector field, $V$, one might propose the DCC-CLRC of Section \ref{sec:lorlatt} with condition (v) modified thus: `the characteristic distance/time over which the continuous geometry $(M,g)$ varies appreciably is everywhere much greater than the Planck length/time \textit{in frames in which $V$ has components} $(V^0, 0 , 0 , 0)$'. Agreed, this would invalidate the counterexample of the gravitational wave burst (\ref{eq:gwburst}), assuming that $V^\mu = (V^0, 0 , 0 , 0)$ in coordinate system $\{t,x,y,z\}$. However, there must be approximate Lorentz invariance in the recovered ATG in order to be consistent with our observations to date 
and it would be necessary to show how the proposed DCC can be compatible with approximate Lorentz invariance.

Or someone might say: `Theory X could be fundamentally discrete, while the Physical Data that is derived from the theory in a grounding state is not Planck scale discrete'. To which, our reply is:  `This is a denial of Assumption 2 as discussed in Comment 3 of Section \ref{subsec:2asspn}. But fair enough: one can reject the rationale for Planck scale spacetime discreteness---including the finite value of the black hole entropy \cite{Sorkin:1985bu}---and maintain that the physics of, and in, a finite spacetime region in General Relativity requires theory X to supply an infinite amount of information; or, at least, information about Lorentzian geometry at sub-Planckian scales. In short: we admit this is possible.' 

\medskip

\noindent {\em Objection 4)}:---`Here is a discrete-continuum correspondence for Combinatorial Complexes that will work: Fill the simplices of the CLRC with patches of flat spacetime so as to form the corresponding Geometric Regge Complex (GRC).  Smooth the corners of this piecewise flat Lorentzian manifold to form the differentiable manifold 
$(M_{CLRC}, g_{CLRC})$. We then declare a GR spacetime $(M,g)$ to be a good approximation to the CLRC if  $(M,g)$ is approximately isometric to $(M_{CLRC}, g_{CLRC})$.' \\
\noindent {\em Reply} :
This natural-seeming proposal for the discrete-continuum correspondence for CLRCs  is based on the assumption that filling in the simplices with patches of flat spacetime is the essentially unique way to produce a spacetime that (i) agrees with the data of a CLRC and (ii) has curvature that only varies appreciably on scales much larger than the Planck scale. In other words, it assumes that if one were to fill in the simplices of a CLRC with non-flat Lorentzian geometry, then the resulting manifold (smoothed if necessary)  would necessarily be unphysical 
because it would have curvature that varies on sub-Planckian scales. 

Now, this assumption may be valid for combinatorial \textit{Riemannian} simplicial complexes (and we believe it is). But it is false for CLRCs: filling in with a non-flat metric does \textit{not} necessarily give a spacetime with curvature that varies on sub-Planckian scale. Indeed, our gravitational wave example  shows exactly that: the spacetime (\ref{eq:gwburst}) is the result of filling in the simplices of $(S, dec)$ with non-flat geometry and yet, (\ref{eq:gwburst}) does \textit{not} have curvature that varies on smaller than Planckian scales as shown by (\ref{eq:gwburstprime}).

\medskip

\noindent {\em Objection 5)}:--- `Your arguments are not background-independent. Both the DCC-C and the DCC-CLRC are given in terms of an embedding of the DPD into a background.' \\
\noindent {\em Reply}: The definitions of DCC-C and DCC-CLRC are perfectly background-independent. For, in neither case is 
$(M,g)$ any kind of background. Rather, $(M,g)$ is the \textit{candidate
recovered spacetime}. The DCC-C for example says: a GR spacetime is recovered by a causal set  if there exists an embedding of the causal set in the spacetime satisfying certain conditions. This is a background-independent and coordinate-independent criterion.  Agreed: we use labels 
and coordinates to define the embeddings that we analyse above. But the existence, or not, of a suitable embedding is a background-independent  condition. 


\section{Causal sets as the basis of a theory of quantum gravity}\label{sec:causal set dynamics}

In this Section we motivate and briefly describe the basics of the causal set programme for quantum gravity (for reviews see \cite{Sorkin:1990bh,Sorkin:1990bj,Sorkin:2003bx,Surya_2019}; and for philosophical discussion see \cite{Wuthrich:2020gwf}). The rationale for doing so in this paper is that it enables us to address a further possible objection to our argument in a concrete setting. The objection is as follows.

\medskip

\noindent\textit{Objection 6}: A quantum gravity theory X will not produce one exact causal set or one exact simplicial complex as DPD. 
There must be some irreducible uncertainty in the data, notwithstanding Remark A of Comment 3 in Section \ref{subsec:2asspn}, viz. that to recover a GR spacetime, one must at some point get classical data. This fundamental uncertainty may (a) disallow causal sets as Planck scale DPD-sets, and-or (b) allow simplicial complexes to overcome the obstacle you have presented in Section \ref{sec:lattices}.

\medskip

We will come to our response shortly. For it will be clearest to set the scene by reviewing the causal set programme. The first thing to do is to develop a little the kinematics/dynamics contrast that we mentioned before, e.g. in Remark A of Section \ref{subsec:2claim}.  Thus it is useful to divide work on causal sets into the categories of kinematics, dynamics and phenomenology: \textit{the substance, its laws, and how it reveals itself}, respectively. The recovery of GR is assumed to be part of the phenomenology of causal set theory if it is to be a successful theory of quantum gravity. And there is also phenomenology that goes beyond the recovery of our known theories. An example of this is the successful prediction of the order of magnitude of the 
cosmological ``constant'' today \cite{Sorkin:1990bj}; and  the further development of the cosmological model of Everpresent Lambda  \cite{Ahmed:2002mj, Zwane:2017xbg}. In this paper, Section  \ref{sec:causalset} has dealt with what one might call the \textit{kinematical emergence} of the 
continuum from causal sets; and so we now address the \textit{dynamical emergence} of the 
continuum from causal sets. 

Causal set theory makes the hypothesis that quantum gravity is based on a path integral, or Sum Over Histories (SOH),  in which the histories summed over are causal sets.  
 The hypothesis that the histories are causal sets brings together
various traditions of thought, including fundamental atomicity, rejection of infinity in physical quantities, and causal order as a more primitive organising principle even than 
 space and time \cite{Robb:1936, Finkelstein:1969}. More modern physical motivations for fundamental spacetime discreteness include the problems of continuum physics  
that Sorkin \cite{Sorkin:1990bh} has dubbed the `three infinities'---$Z = \infty$ (where $Z$ is the partition function,  referring to the infinite values of physical quantities in quantum field theory that renormalisation tries to take care of), $R_{abcd}= \infty$ (referring to the infinite curvatures and tidal forces at singularities predicted by
GR) and 
 $S_{BH} = \infty$ (referring to the infinite entropy of a black hole due to the entanglement of quantum field modes inside and outside the horizon \cite{Sorkin:1985bu}).  
 
 Starting from any one of its main ingredients---the path integral, spacetime discreteness and the primitivity of causal order---there are many roads to causal set theory.  Here is one that begins with the path integral:\\
  (i) The path integral framework for quantum theory
respects the relativistic world view in which the world is fundamentally 4-dimensional, whereas the 
canonical framework does not \cite{Dirac:1933}. We should use the path integral as the basis for a theory of quantum gravity.\\
(ii) Fundamental discreteness of the spacetime histories in the path integral for quantum gravity
eliminates the technical problems of the existence/convergence of the SOH and also has the potential to 
solve the problems of the `three infinities'..\\
 (iii) The KPHM theorem tells us that causal order is the physical information that one needs to add to a discrete manifold  (in Riemann's sense, cf. footnote \ref{Riemann}) of spacetime atoms, in order to recover Lorentzian geometry: for each distinguishing Lorentzian manifold, there are causal sets to which that Lorentzian manifold is a good approximation.\\
  (iv) Overall, a sum over causal sets does justice to the widely-accepted heuristic of a gravitational path integral over geometries, whilst embodying fundamental discreteness. \\

 In the causal set literature, it is 
argued that the SOH is over all causal sets of a fixed cardinality $n$ \cite{Ahmed:2002mj}: let us adopt that assumption. But the set of all causal sets of a fixed cardinality is much larger than the set of causal sets that are faithfully embeddable in 4-dimensional GR spacetimes: for two reasons. (i): The sum includes causal  sets that are faithfully embeddable in any distinguishing Lorentzian manifold of any dimension and any topology (so long as it is consistent with the geometry being slowly varying on the discreteness scale and the spacetime volume being finite) not just 4-dimensional GR spacetimes.
This accords with the broadest construal of the heuristic of the gravitational path integral, namely as being over  all manifolds as well as metrics. (ii): But furthermore the SOH also includes causal sets that are not faithfully embeddable in any Lorentzian geometry at all -- the so-called non-manifold-like causal sets. This class of non-manifold-like causal sets dominates the 
set of all causal sets, in the following sense: the probability that a randomly chosen poset in the set of all posets of cardinality $n$ 
 is a Kleitman-Rothschild (KR) 3-layer poset---i.e. a three-layer poset with roughly $n/2$ elements in the middle layer and $n/4$ elements in the first and third layers---tends to 1, as $n$ tends to infinity \cite{Kleitman:1975}. Moreover the number of KR orders grows like $2^{n^2/4}$, i.e. super-exponentially in $n$. 
 
 This means that the primal struggle between `entropy' and `action' in the SOH for causal sets is of crucial importance.  The KR orders and other non-manifold-like causal sets dominate in number. But they had better not 
 dominate in contribution to the SOH, or there would be no continuum regime: in the terminology of this paper, there would be no  grounding states. This entropic challenge is the manifestation of (one aspect of) the cosmological constant problem in causal set theory. For if the cosmological constant term, $\Lambda$, takes the value expected from local Effective Field Theory `naturalness' arguments, $\Lambda$ would be $10^{120}$ times its observed value and would give rise to curvature on Planckian scales and preclude a smooth continuum regime. 

The action, or more generally, the dynamical laws for causal sets, had better have the wherewithal  to overcome the numerical dominance of non-manifold-like causal sets.  Sorkin has argued that only a \textit{nonlocal} action/dynamics has any hope  of doing this; and this expectation is supported by results in both classical and quantum models for causal set  dynamics. For example, in each model in  the physically motivated class of nonlocal, classical stochastic dynamics known as Classical Sequential Growth, the set of KR orders has probability zero\cite{Rideout:2000a,Rideout:phd}. Another example is in the context of quantum ``state sum'' models: a large class of non-manifold-like causal sets---2-layer orders---is suppressed when weighted by $e^{iS}$  where $S$ is the nonlocal, causal set analogue of the gravitational action  \cite{Loomis:2017jhn}. These are models of the sort of thing that we need. 
  
 On the other hand, our hopes for recovering General Relativity from causal sets depend on our securing locality in the continuum approximation: since this, together with general covariance, gives the Einstein Hilbert term in the familar derivative expansion for the effective field theory in the continuum approximation (as explained on page 523 of \cite{Bombelli:1987aa} and stressed by Sorkin since then, including on page 6. of \cite{Sorkin:1990bj}. 
 
To sum up, there is a Goldilocks scenario in causal set theory:   too much locality and there will be no continuum regime (no solution to the manifestation of the cosmological constant problem in causal set theory); too little locality and there will be no Einstein equations. The
amount of locality in causal set theory needs to be just right. 
  
\medskip

\noindent With this brief review of causal set dynamics in hand, we now turn to the {\em Objection}. We will now assume that a quantum causal set dynamics---of either the state sum variety, or the growth model variety---does suppress the non-manifold-like orders in the SOH, and that there is a continuum regime and grounding states: so that we can address the issue of fundamental uncertainty raised in the Objection. In the context of the SOH for causal sets, the Objection is that the DPD in a grounding state cannot be a single history, a single causal set, since this does not do justice to our expectation of fundamental and irreducible quantum uncertainty. 

\medskip

\noindent {\em Reply}:  We first of all point out that even if the DPD-set were a single, faithfully embeddable
causal set, so that there is indeed no uncertainty about the causal set's structure, there would nevertheless be ``uncertainty'' about the structure of the corresponding continuum approximation on close to Planckian scales. One might say that this is better described as a fundamental \textit{lack} of structure on those small scales. But nevertheless, should the DPD in a grounding state in causal set quantum gravity happen to be one  faithfully embeddable causal set, one could declare that this is how `fundamental quantum uncertainty about spacetime on Planckian scales' turned out to manifest itself in quantum gravity. 

The above point notwithstanding, we admit that we do {\em not} expect causal set quantum gravity to produce grounding states in which the set of DPD is a single faithfully embeddable
causal set. Instead, we expect that a grounding state that recovers a GR spacetime $(M,g)$ will give rise to a DPD-set that is the (affirmation of) the \textit{event}, $E(M,g)$, where $E(M,g)$ is the set of all the causal sets that faithfully embed, at Planckian density, into $(M,g)$.

To justify this expectation and give our promised Reply to
Objection 6, we  need to recall what physical properties are in a path integral quantum theory such as causal set theory. There are two currently existing---and closely related---foundations for quantum theory based fundamentally on the path integral: Hartle's Generalised Quantum Mechanics  (GQM)  \cite{Hartle:2006nx} and Sorkin's Quantum Measure Theory (QMT) \cite{Sorkin:1994dt, Sorkin:2006wq}. In both these frameworks, a physical property takes the form of an \textit{event}: that is, the set of histories which have that property. Thus an event is a subset, $E$, of the set  $\Omega$, which is the set of all the histories in the SOH; and the physical statements we infer from $E$ can be inferred from \textit{any arbitrarily chosen one} of the histories in $E$. The inference thus depends on what is common to the histories in $E$: it does not involve any property of the chosen history that is not shared by all the other histories in $E$. 

We note that Hartle uses the term `coarse-grained history' instead of `event' \cite{Hartle:2006nx}. Both terminologies are useful: `coarse-grained history' emphasises its relationship to the individual fine-grained  histories that are its elements; and `event' signals that path integral quantum theory is a species of measure theory, since `event'  is terminology adopted from stochastic processes and measure theory \cite{Sorkin:2006wq}.  Events are given probabilities or, more generally, measures by the path integral---more precisely, by a double path integral or decoherence functional.\footnote{The main difference between GQM and QMT is that in GQM only events which decohere are considered, whereas in QMT microevents which do not decohere are also considered.}

Though the interpretation of path integral quantum theory, in both GQM and QMT, is work in progress, the identification of  physical properties with events is in both programmes part of the axiomatic foundation. 
Therefore, in causal set theory,  the physical property `spacetime is well-described by the GR spacetime $(M,g)$' is identified with an event, a set of causal sets. And so we arrive at the statement above: `a grounding state that recovers a GR spacetime $(M,g)$ will give rise to a set of DPD that is the (affirmation of) the \textit{event}, $E(M,g)$, where $E(M,g)$ is the set of all of the causal sets that faithfully embed, at Planckian density, into $(M,g)$.' To which we add---to articulate our expectation of fundamental quantum uncertainty---that the grounding state gives rise, as DPD, to no finer event: to no subset of $E(M,g)$. 

So, we respond to facet (a) of the ``uncertainty'' Objection, as follows.  For the DPD to be, not a single causal set, but an event, $E(M,g)$, fulfils the causal set programme's expectation of fundamental, quantum uncertainty in quantum gravity. The elements of $E(M,g)$ have the common properties that (i) they are each approximated by $(M,g)$ and (ii) they are each discrete at the Planck scale and have no structure at sub-Planckian scales. That is, both these properties, (i) and (ii), are properties of each  causal set in $E(M,g)$. We can then, \textit{for practical purposes of recovery of} $(M,g)$ take any one of the causal sets in $E(M,g)$---it matters not which one---as the DPD-set that we store as bits in our classical computer, and that recovers $(M,g)$. 

Therefore, not only is fundamental quantum uncertainty not a valid objection to our argument; but also we see explicitly how there need be no contradiction between there being fundamental quantum uncertainty and a single causal set being the DPD that recovers $(M,g)$. 

Turning to facet (b)  of the ``uncertainty'' Objection: consider a putative path integral theory of quantum gravity in which the histories are CLRCs, as we defined in Section \ref{sec:lorlatt}. Consider an event $E$, i.e. a set of CLRCs.
No matter how we choose event $E$, not one of the CLRCs in $E$ recovers a GR spacetime. So there can certainly be no common property of the histories of $E$ that is a continuum approximation.
Fundamental uncertainty of the form considered here, therefore, does not allow CLRCs to recover 
GR spacetimes. Indeed it makes matters worse. A single CLRC cannot recover a GR spacetime because it 
lacks information about the geometry in the voids. The information in an event is a \textit{coarse-graining} of the information in the individual histories in the event; and so the event contains \textit{less} information than each of its individual members.\footnote{Closely related arguments are made, with an emphasis on semi-classical measurements, in \cite{Henson:2009yb}.}\\

A final remark {\em a propos} this Objection: we emphasise again that this paper is \textit{not} claiming that there are grounding states and a continuum regime in quantum gravity based on a SOH over causal sets. Rather, we claim that, if there are such grounding states, then causal sets can do the job of recovering GR spacetimes and being Planck scale DPD in a way that is compatible with fundamental quantum uncertainty.

\section{Conclusion}\label{sec:concl}

The unity of physics demands that before giving up on 
so much scientific progress already made, 
we should take seriously the aim of recovering General Relativity,  
with its Lorentzian geometry and Lorentz invariance,  from quantum gravity.  
We have argued that at our current state of knowledge, a causal set is the 
only kind of entity that can be discrete at the Planck 
scale and can adequately encode the geometric information in 
a Lorentzian spacetime at macroscopic scales.

Further, if one accepts our justification for the two Claims (given in Sections \ref{sec:causalset} and \ref{sec:lattices}), then it is tempting to  conjecture that no challenger to causal sets will arise in the future. More precisely, it is tempting to conjecture that a causal set is the unique Planck-scale DPD-set that can recover a GR spacetime as a continuum approximation. If that is indeed so, the stronger is the motivation for an approach to quantum gravity that is based fundamentally on causal sets. 

\section{Acknowledgments}

We thank  David Meyer, David Rideout, Rafael Sorkin, Sumati Surya and Yasaman Yazdi for helpful discussions.  FD and JB acknowledge the support of 
 the Leverhulme/Royal Society interdisciplinary  APEX grant APX/R1/180098. FD
is supported in part by Perimeter Institute for Theoretical Physics. Research at Perimeter Institute is supported by the Government of Canada through
Industry Canada and by the Province of Ontario through the Ministry of Economic Development and Innovation. FD is supported in part by STFC grant
ST/P000762/1. 

  \appendix
 \section{Appendix}\label{appendix}
 
We assume the function $h(u) $ satisfies the condition 
\begin{equation}\label{vanish}
\int_0^\frac{1}{\sqrt{2}} du \, h(u) = 0\,.
\end{equation}
 
No embedded vertices of the complex lie in the support of $h(u)$ (see figure
\ref{figure1}). There are two coordinate-hypercubes that intersect the
support of $h(u)$ in different ways, see figure \ref{figure1}. Hypercube I
is the cube located at the origin $(0,0,0,0)$ and hypercube II is located at
 $(1,0,0,0)$. 
Each of the other coordinate-hypercubes that intersect the support of $h(u)$
is isometric to one of these two. 

To first order in $\epsilon$, the geodesic equations for  $(u(\lambda),
v(\lambda), x(\lambda), y(\lambda))$, with $\lambda$ an affine parameter,
are as follows; cf. eq. \ref{eq:gwburst}. 
\begin{align*}
\dot{x}& = c_x ( 1- \epsilon h(u) )\,, \ \ \textrm{where}\ c_x \ \textit{is
a constant}\,,\\
\dot{y}& = c_y ( 1+ \epsilon h(u) )\,, \ \ \textrm{where}\ c_y \ \textit{is
a constant}\,,\\
\dot{u}& = c_u \,, \ \ \textrm{where}\ c_u \ \textit{is a constant}\,,\\
\zeta & = - 2\dot{v} c_u + (1-\epsilon h(u) ) c_x^2 +   (1+\epsilon h(u) )
c_y^2\,,
\end{align*}
where dot denotes derivative w.r.t. $\lambda$ and $\zeta <0$, $\zeta=0$ or $\zeta>0$
depending on whether 
the geodesic is timelike, null or spacelike, respectively. When $c_u \ne 0$,
$u$ is also an affine parameter. \\

\noindent{\bf{Hypercube I}} Consider the 15 edge-geodesics in hypercube I
that start at $(0,0,0,0)$. Since the initial value of $u$ is zero and the 
derivative of $u$ along a geodesic is constant, $u$ must either remain equal to  $0$ or increase steadily along the 
geodesic reaching its final value $\frac{1}{\sqrt{2}}$ at the other vertex.
Only in the latter case does the geodesic enter the support of $h(u)$.  
Therefore, all embedded edges except those labelled by lattice vectors
8,10,12 and 14 are outside the 
support of $h(u)$ and they are therefore straight lines in the coordinates
$\{t,x,y,z\}$ and have the same lengths as they do in Minkowski space. We
consider each of the remaining edges in turn and give the geodesic 
to first order in $\epsilon$. In all cases, the coordinate $u$ is an affine
parameter and increases from 0 to $\frac{1}{\sqrt{2}}$ along the geodesic. 
\begin{itemize}
\item[] $8 = (1,0,0,0)$:  The geodesic is timelike and is the straight line
at $x=y=z=0$ 
of proper time duration 1.  
\item[]$10 = (1,0,1,0)$:  The geodesic is null and is given by,
\begin{align*} 
\frac{dt}{du} &= \frac{1}{\sqrt{2}} (2 + \epsilon h(u) ) \\
\frac{dx}{du} &= 0 \\
\frac{dy}{du} &= \sqrt{2} ( 1 + \epsilon h(u) ) \\
\frac{dz}{du} & =  \frac{1}{\sqrt{2}} \epsilon h(u)  \,.
\end{align*} 
\item[]$12 = (1,1,0,0)$:   This is similar to 10. The geodesic is null and
is given  by,
\begin{align*} 
\frac{dt}{du} &= \frac{1}{\sqrt{2}} (2 - \epsilon h(u) ) \\
\frac{dx}{du} &=  \sqrt{2} ( 1 - \epsilon h(u) )  \\
\frac{dy}{du} &= 0 \\
\frac{dz}{du} & =  - \frac{1}{\sqrt{2}} \epsilon h(u)  \,.
\end{align*} 
\item[]$14 = (1,1,1,0)$:  The geodesic is spacelike has proper length 1 and
is given by,
\begin{align*} 
\frac{dt}{du} &= \sqrt{2} \\
\frac{dx}{du} &=  \sqrt{2} ( 1 - \epsilon h(u) )  \\
\frac{dy}{du} &= \sqrt{2} ( 1 + \epsilon h(u) )  \\
\frac{dz}{du} & =0 \,.
\end{align*} 
\end{itemize}

 \noindent{\bf{Hypercube II}} There are 15 edge-geodesics in hypercube II
that start at $(1,0,0,0)$, labelled by the lattice vectors 1-15.  Of those
edges all except 1,3,5 and 7 are outside the 
support of $h(u)$ and are straight lines in the coordinates $\{t,x,y,z\}$
and have the same lengths as they do in Minkowski space. We list the
remaining geodesics 
to first order in $\epsilon$. In all cases, the coordinate $u$ is an affine
parameter and decreases from $\frac{1}{\sqrt{2}}$ to $0$ along the geodesic.
\begin{itemize}
\item[] $1 = (0,0,0,1)$:  The geodesic is spacelike and is the coordinate straight line from $(1,0,0,0)$ to $(1,0,0,1)$. It has proper length 1.  
\item[]$3 = (0,0,1,1)$:  The geodesic is spacelike, is of proper length
$\sqrt{2}$ and is given by,
\begin{align*} 
\frac{dt}{du} &=  \frac{1}{\sqrt{2}} \epsilon h(u) \\
\frac{dx}{du} &= 0 \\
\frac{dy}{du} &= - \sqrt{2} ( 1 + \epsilon h(u) ) \\
\frac{dz}{du} & = - \frac{1}{\sqrt{2}}(2 - \epsilon h(u) )  \,.
\end{align*} 
\item[]$5 = (0,1,0,1)$:   This is similar to 3. The geodesic is spacelike,
is of proper length $\sqrt{2}$ and is given by,
\begin{align*} 
\frac{dt}{du} &= - \frac{1}{\sqrt{2}} \epsilon h(u) \\
\frac{dx}{du} &= - \sqrt{2} ( 1 - \epsilon h(u) )  \\
\frac{dy}{du} &= 0 \\
\frac{dz}{du} & =-\frac{1}{\sqrt{2}}(2 + \epsilon h(u) )  \,.
\end{align*} 
\item[]$7 = (0,1,1,1)$:  The geodesic is spacelike,  is 
the coordinate straight line from $(1,0,0,0)$ to $(1,1,1,1)$ and has proper length  $\sqrt{3}$. 
\end{itemize}

In summary, all the edge-geodesics between the embedded vertices with
integer coordinates have the same lengths as they 
do in Minkowski spacetime, to first order in $\epsilon$.


\bibliography{../Bibliography/refs_from_jeremy_1_Jun}
\bibliographystyle{../Bibliography/JHEP}

\end{document}